\documentclass[12pt]{article}
\usepackage{euscript,amsmath, amssymb, amsfonts}
\usepackage{color}

\newcommand{ \supp}{{\rm supp\,}}
\newcommand{\R}{{\mathbb R}}

\newcommand{\p}{{\hbar^2}}
\newcommand{\pp}[2]{\hbar^{#1}}
\newcommand{\III}{{\EuScript I}}
\newcommand{\ttt}{{\tau}}
\newcommand{\oC}{{\mathbb C}}
\newcommand{\reve}[1]{{\mathbf #1}}

\newcommand{\di}{{ d }}
\newcommand{\D}{\EuScript D}
\newcommand{\eE}{\EuScript E}

\newcommand{\G}{{\mathbb G}}
\newcommand{\Z}{{\mathbb Z}}
\newcommand{\K}{{\mathbb K}}
\newcommand{\be}{\begin{equation}}
\newcommand{\ee}{\end{equation}}

\newtheorem{theorem}{Theorem}

\newtheorem{definition}[theorem]{Definition}

\newtheorem{proposition}[theorem]{Proposition}
\newenvironment{proof}[1][Proof]{\noindent\textsf{#1.} }{\ \rule{0.5em}{0.5em}}

\font\frtnfr=eufm10 scaled\magstep1
\font\twlfr=eufm10
\font\tenfr=eufm10

\newfam\frfam
\textfont\frfam=\frtnfr
\scriptfont\frfam=\twlfr
\scriptscriptfont\frfam=\tenfr

\font\frtnopen=msbm10 scaled\magstep2
\font\twlopen=msbm10
\font\tenopen=msbm10

\newfam\openfam
\textfont\openfam=\frtnopen
\scriptfont\openfam=\twlopen
\scriptscriptfont\openfam=\tenopen

\font\frtnsf = cmss12 scaled\magstep1
\font\twlsf = cmss10
\font\tensf = cmss9

\newfam\Scfam
\textfont\Scfam = \frtnsf
\scriptfont\Scfam = \twlsf
\scriptscriptfont\Scfam = \tensf

\begin{document}

\sloppy \title { Deformations of the central extension of \\the
 Poisson superalgebra. } \author {
 S.E.Konstein\thanks{E-mail: konstein@lpi.ru}\ \ and
 I.V.Tyutin\thanks{E-mail: tyutin@lpi.ru} \thanks{ This work was
 supported by the R{}FBR (grants No.~02-02-16944 (I.T.)
 and No.~02-02-17067 (S.K.)), and by the grant
 LSS-1578.2003.2. } \\ {\sf \small I.E.Tamm Department
 of Theoretical Physics,} \\ {\sf \small P. N. Lebedev
 Physical Institute,} \\ {\sf \small 119991, Leninsky
 Prospect 53, Moscow, Russia.} } \date { }

\maketitle

\begin{abstract} { \footnotesize Poisson superalgebras realized on
the smooth Grassmann valued functions with compact support in $\R^n$
have the central extensions. The deformations of these central
extensions are found. } \end{abstract}

%%%%%%%%%%%%%%%%%%%%%%%%%%%%%%%%%%%%%%%%%%%%%%%%%%%

\section{Introduction}

This paper continues the investigation started in \cite{n>2},
\cite{deform} and \cite{n=2}. We consider the Poisson superalgebra
realized on smooth Grassmann-valued functions with compact support in
${\mathbb{R}}^{n}.$ As it is shown in \cite{n>2} this superalgebra
has central extensions for some dimensions. {}For these dimensions,
we found the second adjoint cohomology space and the deformations of
the Poisson superalgebra under consideration.

\subsection{Poisson superalgebra} Let $\K$ be either $\R$ or $\oC$.
We denote by $\EuScript D(\R^n)$ the space of smooth $\K$-valued
functions with compact support on $\R^n$. This space is endowed with
its standard topology. We set $$ D^{n_-}_{n_+}= \EuScript
D(\R^{n_+})\otimes \G^{n_-},\quad E^{n_-}_{n_+}=
C^\infty(\R^{n_+})\otimes \G^{n_-},\quad D^{\prime n_-}_{n_+}=
\EuScript D'(\R^{n_+})\otimes \G^{n_-}, $$ where $\G^{n_-}$ is the
Grassmann algebra with $n_-$ generators and $\EuScript D'(\R^{n_+})$
is the space of continuous linear functionals on $\EuScript
D(\R^{n_+})$. The generators of the Grassmann algebra (resp., the
coordinates of the space $\R^{n_+}$) are denoted by $\xi^\alpha$,
$\alpha=1,\ldots,n_-$ (resp., $x^i$, $i=1,\ldots, n_+$). We shall
also use collective variables $z^A$ which are equal to $x^A$ for
$A=1,\ldots,n_+$ and are equal to $\xi^{A-n_+}$ for
$A=n_++1,\ldots,n_++n_-$. The spaces $ D^{n_-}_{n_+}$, $
E^{n_-}_{n_+}$, and $ D^{\prime n_-}_{n_+}$ possess a natural grading
which is determined by that of the Grassmann algebra. The parity of
an element $f$ of these spaces is denoted by $\varepsilon(f)$. We
also set $\varepsilon_A=0$ for $A=1,\ldots, n_+$ and
$\varepsilon_A=1$ for $A=n_++1,\ldots, n_++n_-$. Each function $f\in
E^{n_-}_{n_+}$ can be expressed in the form
$$f(z)=\sum_{\alpha_i=0,1} f_{\alpha_1,...,\alpha_{n_-}}(x)
\xi_1^{\alpha_1} ...\xi_{n_-}^{\alpha_{n_-}}.$$ We will use the
notation $\mathrm{supp}(f)=\bigcup_{\{\alpha_i\}}
\mathrm{supp}(f_{\alpha_1,...,\alpha_{n_-}}) \subset \R^{n_+}$.

Let $\partial/\partial z^A$ and $\overleftarrow{\partial}/\partial
z^A$ be the operators of the left and right differentiation. The
Poisson bracket is defined by the relation \begin{equation}
\{f,g\}(z)=f(z)\frac{\overleftarrow{\partial}}{\partial
z^A}\omega^{AB} \frac{\partial}{\partial z^B}g(z)=
 -(-1)^{\varepsilon(f)\varepsilon(g)}\{g,f\}(z), \label{3.0}
\end{equation} where the symplectic metric
$\omega^{AB}=-(-1)^{\varepsilon_A\varepsilon_B} \omega^{BA}$ is a
constant invertible matrix. {}For definiteness, we choose it in the
form $$ \omega^{AB}= \left(\begin{array}{cc}\omega^{ij}&0 \\
0&\lambda_\alpha\delta^{\alpha\beta}\end{array}\right),\quad
\lambda_\alpha=\pm1,\ i,j=1,...,n_+,\ \alpha,\beta=1+n_+,...,n_-+n_+,
$$ where $\omega^{ij}$ is the canonical symplectic form (if $\K=\oC$,
then one can choose $\lambda_\alpha=1$). The nondegeneracy of the
matrix $\omega^{AB}$ implies, in particular, that $n_+$ is even. The
Poisson superbracket satisfies the Jacobi identity \begin{equation}
(-1)^{\varepsilon(f)\varepsilon(h)}
\{f,\{g,h\}\}(z)+\hbox{cycle}(f,g,h)= 0,\quad f,g,h\in E^{n_-}_{n_+}.
\label{2.3.1a} \end{equation}

By Poisson superalgebra, we mean the space $ D^{n_-}_{n_+}$ with the
Poisson bracket~(\ref{3.0}) on it. The relations~(\ref{3.0})
and~(\ref{2.3.1a}) show that this bracket indeed determines a Lie
superalgebra structure on $ D^{n_-}_{n_+}$.

The integral on $ D^{n_-}_{n_+}$ is defined by the relation $\bar f
\stackrel {def} = \int \di z\, f(z)= \int_{\R^{n_+}}\di x\int
\di\xi\, f(z)$, where the integral on the Grassmann algebra is normed
by the condition $\int \di\xi\, \xi^1\ldots\xi^{n_-}=1$. We identify
$\G^{n_-}$ with its dual space $\G^{\prime n_-}$ setting
$f(g)=\int\di\xi\, f(\xi)g(\xi)$, $f,g\in \G^{n_-}$.
Correspondingly, the space $ D^{\prime n_-}_{n_+}$ of continuous
linear functionals on $ D^{n_-}_{n_+}$ is identified with the space
$\D^\prime(\R^{n_+})\otimes \G^{n_-}$. As a rule, the value $m(f)$
of a functional $m\in D^{\prime n_-}_{n_+}$ on a test function $f\in
D^{n_-}_{n_+}$ will be written in the ``integral'' form: $ m(f)
=\int \di z\, m(z) f(z).$

\subsection{Cohomologies of Poisson superalgebra} Let $L$ be a Lie
superalgebra acting in a $\Z_2$-graded space $V$ (the action of
$f\in L$ on $v\in V$ will be denoted by $f\cdot v$). The space $C_
p(L, V)$ of $p$-cochains consists of all multilinear
super-antisymmetric mappings from $L^p$ to $V$. The space $C_p(L,
V)$ possesses a natural $\Z_2$-grading: by definition, $M_p\in C_
p(L, V)$ has the definite parity $\varepsilon(M_p)$ if $$
\varepsilon(M_p(f_1,\ldots,f_p))= \varepsilon(M_p)+\varepsilon(f_
1)+\ldots+\varepsilon(f_1) $$ for any $f_j\in L$ with parities
$\epsilon(f_j)$. The differential $\di_p^V$ is defined to be the
linear operator from $C_p(L, V)$ to $C_{p+1}(L, V)$ such that
\begin{align} &&d_p^{V}M_p(f_1,...,f_{p+1})= -\sum_
{j=1}^{p+1}(-1)^{j+\varepsilon(f_j)|\varepsilon(f)|_{1,j-1}+
\varepsilon(f_j)\varepsilon_{M_p}}f_j\cdot M_p(f_
{1},...,\breve{f}_j,...,f_{p+1})- \nonumber \\ &&-\sum_
{i<j}(-1)^{j+\varepsilon(f_j)|\varepsilon(f)|_{i+1,j-1}} M_p(f_
1,...f_{i-1},\{f_i,f_j\},f_{i+1},..., \breve{f}_j,...,f_{p+1}),
\label{diff} \end{align} for any $M_p\in C_p(L, V)$ and $f_
1,\ldots f_{p+1}\in L$ having definite parities. Here the sign
$\breve{}$ means that the argument is omitted and the notation $$
|\varepsilon(f)|_{i,j}=\sum_{l=i}^j\varepsilon(f_l) $$ has been
used. The differential $\di^V$ is nilpotent (see \cite{Schei97}),
{\it i.e.}, $\di^V_{p+1}\di^V_p=0$ for any $p=0,1,\ldots$. The
$p$-th cohomology space of the differential $\di_p^V$ will be
denoted by $H^p_V$. The second cohomology space $H^2_{\mathrm{ad}}$
in the adjoint representation is closely related to the problem of
finding formal deformations of the Lie bracket $\{\cdot,\cdot\}$ of
the form $ \{f,g\}_*=\{f,g\}+\hbar^2 \{f,g\}_1+\ldots. $ The
condition that $\{\cdot,\cdot\}$ is a 2-cocycle is equivalent to the
Jacobi identity for $\{\cdot,\cdot\}_*$ modulo the $\hbar^4$-order
terms.

In \cite{n>2} and \cite{n=2}, we studied the cohomologies of the
Poisson algebra $ D^{n_-}_{n_+}$ in the following representations:
\begin{enumerate} \item The trivial representation: $V=\K$, $f\cdot
a=0$ for any $f\in D^{n_-}_{n_+}$ and $a\in\K$. The space $C_p(
D^{n_-}_{n_+}, \K)$ consists of separately continuous
antisymmetric multilinear forms on $( D^{n_-}_{n_+})^p$. The
cohomology spaces and the differentials is denoted by $H^p_
{\mathrm{tr}}$ and $\di^{\mathrm{tr}}_p$ respectively. \item The
adjoint representation: $V=D^{n_-}_{n_+}$ and $f\cdot g=\{f,g\}$
for any $f,g\in D^{n_-}_{n_+}$. The space $C_p( D^{n_-}_{n_
+}, D^{n_-}_{n_+})$ consists of separately continuous
antisymmetric multilinear mappings from $(D^{n_-}_{n_+})^p$ to $
D^{n_-}_{n_+}$. The cohomology spaces and the differentials is
denoted by $H^p_{\mathrm{ad}}$ and $\di^{\mathrm{ad}}_p$
respectively. \end{enumerate}

The following theorems are proved in \cite{n>2} and \cite{n=2}:

\begin{theorem} $ $ \label{th1}

Let bilinear forms $C_2^1$ and $C_2^2$ be defined by the relations
\begin{eqnarray} C_2^1(f,g)=\bar f\bar g, \\ C_2^2(f,g)= \int
 dz\left(f(z){}\eE_z g(z)- (-1)^{\varepsilon(f)\varepsilon(g)}
g(z){}\eE_z f(z)\right),\quad f,g\in D^{n_-}_{n_+}, \label{c22}
\end{eqnarray} where $\eE_z \stackrel {def} = 1-\frac 1 2 z^A \frac
{\partial} {\partial z^A}.\,$ \footnote{The operator ${\cal E}_z$
is a derivation of the Poisson superalgebra.}

If $n_-$ is even and $n_- \ne n_+ + 4$, then $H^2_
{\mathrm{tr}}=0$;

if $n_- = n_+ +4$, then $H^2_{\mathrm{tr}}\simeq \K$ and the form
$C^2_2$ is a nontrivial cocycle;

if $n_-$ is odd, then $H^2_{\mathrm{tr}}\simeq \K$ and the form
$C^1_2$ is a nontrivial cocycle. \end{theorem}

\begin{theorem} $ $ \label{th0} Let $V_1$ be the one-dimensional
subspace of $C_1(\mathbf D^{n_-}_{n_+},\mathbf D^{n_-}_{n_+})$
generated by the cocycle $$m_{1|2}=\eE_z f(z).$$ Then there is a
natural isomorphism $V_1\oplus (\mathbf E^{n_-}_{n_+}/\mathbf
D^{n_-}_{n_+})\simeq H^1_{\mathrm{ad}}$ taking $(M_1,T)\in V_
1\oplus (\mathbf E^{n_-}_{n_+}/\mathbf D^{n_-}_{n_+})$ to the
cohomology class determined by the cocycle $M_1(z|f)+\{t(z),f(z)\}$,
where $t\in \mathbf E^{n_-}_{n_+}$ belongs to the equivalence
class $T$. \end{theorem}

\begin{theorem} $ $ \label{th2}

Let $n_+\geq 4$ or ($n_+=2$ and $n_-\ge 4$)

Let $V_2$ be the subspace of $C_2(D^{n_-}_{n_+},D^{n_-}_{n_
+})$ generated by the bilinear mappings $m_{2|1}$ and $m_{2|3}$
from $(D^{n_-}_{n_+})^2$ to $ D^{n_-}_{n_+}$, which are defined
by the relations \begin{eqnarray} && m_
{2|1}(z|f,g)=f(z)\!\left(\frac{\overleftarrow{\partial}}{\partial
z^A} \omega^{AB}\frac{\partial}{\partial z^B}\right)^3\!g(z),
\label{2|1} \\ && m_{2|3}(z|f,g)= \bar g \eE_z f-
(-1)^{\varepsilon(f)\varepsilon(g)} \bar f \eE_z g. \label{2|3}
\end{eqnarray}

Then the cochains $m_{2|1}$ and $m_{2|3}$ are independent
nontrivial cocycles, $dim V_2=2$, and there is a natural isomorphism
$V_2\oplus (E^{n_-}_{n_+}/D^{n_-}_{n_+})\simeq H^2_
{\mathrm{ad}}$ taking $(M_2,T)\in V_2\oplus (E^{n_-}_{n_+}/D^{n_
-}_{n_+})$ to the cohomology class determined by the cocycle $$ M_
2(z|f,g)-\{t(z),f(z)\}\bar{g}+
(-1)^{\varepsilon(f)\varepsilon(g)}\{t(z),g(z)\}\bar{f}, $$ where $t
\in E^{n_-}_{n_+}$ belongs to the equivalence class $T$.
\end{theorem}

\begin{theorem} $ $ \label{th3} \begin{enumerate} \item
\label{item1.3} Let $n_+=2$, $0 \le n_- \le 3$. Let $N_1(f) =
-2\Lambda(x_2)\int du \theta(x_1-y_1)f(u),$ \noindent where
$\Lambda\in C^\infty({\mathbb R})$ be a function such that $\frac d
{dx} \Lambda \in \D(\R)$ and $\Lambda(-\infty) = 0,\ \Lambda( +
\infty) = 1$. Let \begin{eqnarray} z &=& (x_1,x_2,\xi_
1,\,...\,,\,\xi_{n_-}),\ \ \ u = (y_1,y_2,\eta_1,\,...\,,\,\eta_
{n_-}) \notag\\ \Theta(z|f) & =& \int du \delta(x_1-y_1)\theta(x_
2-y_2)f(u), \nonumber\\ N^E_2(z|f,g) &=&
\Theta(z|\partial_{2}fg)-\Theta(z|f\partial_{2}g)-2(-1)^{n_
-\varepsilon (f)} \partial_{2}f(z)\Theta(z|g) +
2\Theta(z|f)\partial_{2}g(z), \notag\\ N^D_2(f,g) &=& N^E_
2(f,g) + d_1^{\mathrm{ad}}N_1(f,g),\\ \Delta (z|f) &= &\int du
\delta(x-y) f(u), \\ \Delta _{\alpha}(z|f) & = &\int du \eta_
\alpha\delta(x-y) f(u). \end{eqnarray}

Then the bilinear mappings $L_2^{n_-}$ from $(D^{n_-}_{2})^2$ to
$E^{n_-}_{2}$ defined by the relations \begin{eqnarray} L^0_
2(x|f,g)& = & N^D_2(x|f,g) + \frac{1}{2} \left ( x^{i} \partial_
{i}f(x) \right) g(x) -\frac{1}{2} f(x) \left ( x^i\partial_
{i}g(x)\right), \nonumber\\ L^1_2(z|f,g)& = & N^D_2(z|f,g)- \Delta
(z|f) g(z) + (-1)^{\varepsilon(f)} f(z)\Delta (z|g) \nonumber \\ &&
-\frac{2}{3} (-1)^{\varepsilon(f)} \left ( \xi^1\partial_{\xi^1}f(z)
\right) \Delta (z|g), \label{L21} \\ L^2_2(z|f,g)& = &N^D_
2(z|f,g)-\Delta (z|f)g(z) + f(z)\Delta(z|g) \nonumber\\ L^3_
2(z|f,g)& = &N^D_2(z|f,g) -\Delta (z|f)g(z)+ (-1)^{\varepsilon
(f)}f(z)\Delta (z|g) + \nonumber\\ && +
\partial _{\xi ^{\alpha }}f(z)\Delta _{\alpha }(z|g)-
(-1)^{\varepsilon (f)} \Delta _{\alpha}(z|f)\partial _{\xi ^{\alpha
}}g(z), \nonumber \end{eqnarray} are cocycles and maps $(D^{n_-}_
{2})^2$ to $D^{n_-}_{2}$.

\item \label{item1.4} Let $n_+=2$ and $0\le n_-\le 3$. Let $V^{n_
-}_2$ be the subspace of $C_2(D^{n_-}_{2},D^{n_-}_{2})$
generated by the cocycles $M^1_2$, $M^3_2$ and $L_2^{n_-}$, where
the cocycles $M^1_2$ and $M^3_2$ are defined in Theorem \ref{th2}.

Then there is a natural isomorphism $V^{n_-}_2\oplus (E^{n_-}_
{2}/ D^{n_-}_{2}) \simeq H^2_{\mathrm{ad}}$ taking $(M_2,T)\in
V^{n_-}_2\oplus (E^{n_-}_{2}/D^{n_-}_{2})$ to the cohomology
class determined by the cocycle $$ M_2(z|f,g)-\{t(z),f(z)\} \bar{g}
+ (-1)^{\varepsilon(f)\varepsilon(g)}\{t(z),g(z)\} \bar{f}, $$ where
$t\in E^{n_-}_{2}$ belongs to the equivalence class $T$.
\end{enumerate} \end{theorem}

\subsection{Central extensions} Let $\reve{L}$ be the central
extension of the superalgebra $L$, i.e. $\reve{L} =L\oplus c$,
$\reve{f}=f+a \III \in \reve{L}$, $a \III \in c$, $\varepsilon ( \III
)=0$, $a\in \K$. The bracket in $\reve{L}$ we denote as
$[\reve{f},\reve{g}]= -(-1)^{\varepsilon (\reve{f})\varepsilon
(\reve{g})}[\reve{g},\reve{f}] $,

\begin{equation*} [ f,g]=\left\{ f,g\right\} +C(f,g) \III
,\;C(f,g)\in \K,\;[\reve{f} , \III ]=0, \end{equation*} where
$C(f,g)=-(-1)^{\varepsilon (f)\varepsilon (g)}C(g,f)$, $\varepsilon
(C)=0$, is a generator of the second cohomology in the trivial
representation of the algebra $L$: \begin{equation*} d_
{2}^{\mathrm{tr}}C(f,g,h)=C(\left\{ f,g\right\} ,h)-(-1)^{\varepsilon
(g)\varepsilon (h)}C(\left\{ f,h\right\} ,g)-C(f,\left\{ g,h\right\}
)=0. \end{equation*}

It follows from Theorem \ref{th1}, that the Poisson superalgebra
$D^{n_-}_{n_+}$ has the central extension $\mathbf D^{n_-}_{n_
+}$ either if $n_-$ is odd or if $n_-=n_+ +4$.

Consider $C_p(\reve L,\reve L)$.

If $\reve M_p\in C_p(\reve L,\reve L)$ then $ \reve{M}_
{p}(\reve{f},...)=M_{p}(\reve{f},...)+m_{p}(\reve{f},...)\III \in
\reve{L}, $ where $M_{p}\in C_p(\reve L, L)$, $m_{p}\in C_p(\reve
L,\K)$, $\varepsilon (\reve{M} _{p})=\varepsilon (M_
{p})=\varepsilon (m_{p})=\varepsilon _{M_{p}}$.

The differential $\reve{d}^{\mathrm{ad}}$ is defined in a usual way.

{}For linear forms, the differential has the form: \begin{equation*}
\reve{d}_{1}^{\mathrm{ad}}\reve{M}_
{1}(\reve{f},\reve{g})=[\reve{M}_{1}(
\reve{f}),\reve{g}]-(-1)^{\varepsilon (\reve{f})\varepsilon
(\reve{g})}[ \reve{M}_{1}(\reve{g}),\reve{f}]-\reve{M}_
{1}([\reve{f},\reve{g} ]) \end{equation*} and can be expressed on the
decomposition of $\reve L$ as \begin{eqnarray} && \reve{d}_
{1}^{\mathrm{ad}} \reve{M}_{1}(\III ,\III )\equiv 0\\ &&\reve{d}_
{1}^{\mathrm{ad}}\reve{M}_{1}(f,\III )= -\left\{M_{1}(\III
),f\right\} -C(M_{1}(\III ),f)\III , \\ && \reve{d}_
{1}^{\mathrm{ad}}\reve{M}_{1}(f,g)= d_{1}^{\mathrm{ad}}M_
{1}(f,g)-M_{1}(\III )C(f,g)+\gamma (f,g)\III , \end{eqnarray} where
\be \gamma (f,g)=C(M_{1}(f),g)-(-1)^{\varepsilon (f)\varepsilon
(g)}C(M_{1}(g),f)-m_{1}(\left\{ f,g\right\} )-m_{1}(\III )C(f,g).
\ee

{}For bilinear forms, the differential has the form:

\begin{eqnarray*} \reve{d}_{2}^{\mathrm{ad}}\reve{M}_
{2}(\reve{f},\reve{g},\reve{h}) &=&(-1)^{\varepsilon
(\reve{g})\varepsilon (\reve{h})} [\reve{M}_{2}(
\reve{f},\reve{h}),\reve{g}]- [\reve{M}_
{2}(\reve{f},\reve{g}),\reve{h} ]- \\ &&\,-(-1)^{\varepsilon
(\reve{f})\varepsilon (\reve{g})+ \varepsilon (\reve{f})\varepsilon
(\reve{h})} [\reve{M}_{2}(\reve{g},\reve{h}),\reve{f}]- \\
&&\,-\reve{M}_{2}([\reve{f},\reve{g}],\reve{h})+\;(-1)^{\varepsilon
( \reve{g})\varepsilon (\reve{h})}\reve{M}_
{2}([\reve{f},\reve{h}],\reve{ g})+\reve{M}_
{2}(\reve{f},[\reve{g},\reve{h}]) \end{eqnarray*} and can be
expressed on the decomposition of $\reve L$ as \begin{eqnarray}
&&\!\!\!\!\!\!\!\!\! \reve{d}_{2}^{\mathrm{ad}}\reve{M}_
{2}(f,g,h)=d_{2}^{\mathrm{ad} }M_{2}(f,g,h)+ \notag\\ &&+[M_
{2}(f,\III )C(g,h)-(-1)^{\varepsilon (f)\varepsilon (g)}M_{2}(g,\III
)C(f,h) +(-1)^{\varepsilon (f)\varepsilon (h)+\varepsilon
(g)\varepsilon (h)}M_{2}(h,\III )C(f,g)]+ \notag\\ &&
+[(-1)^{\varepsilon (g)\varepsilon (h)}C(M_{2}(f,h),g)-C(M_
{2}(f,g),h) -(-1)^{\varepsilon (f)\varepsilon (g)+\varepsilon
(f)\varepsilon (h)}C(M_{2}(g,h),f)+ \notag\\ && +m_{2}(f,\III
)C(g,h)-(-1)^{\varepsilon (f)\varepsilon (g)}m_{2}(g,\III )C(f,h)
+(-1)^{\varepsilon (f)\varepsilon (h)+\varepsilon (g)\varepsilon
(h)}m_{2}(h,\III )C(f,g)- \notag\\ && -d_{2}^{\mathrm{tr}}m_
{2}(f,g,h)]\III , \label{1.2} \end{eqnarray} \begin{eqnarray}
&&\!\!\!\!\!\!\!\!\! \reve{d}_{2}^{\mathrm{ad}}\reve{M}_
{2}(f,g,\III )=d_{1}^{\mathrm{ad}} \tilde{M}_{1}(f,g) +[C(M_
{2}(f,\III ),g)-(-1)^{\varepsilon (f)\varepsilon (g)}C(M_{2}(g,\III
),f)- \notag\\ &&-m_{2}(\left\{ f,g\right\} ,\III )]\III ,
\label{1.3} \\ &&\!\!\!\!\!\!\!\!\! \reve{d}_
{2}^{\mathrm{ad}}\reve{M}_{2}(f,\III ,\III )=\reve{d}_{2}^{
\mathrm{ad}}\reve{M}_{2}(\III ,\III ,\III )\equiv 0, \end{eqnarray}
where \begin{equation*} \tilde{M}_{1}(f)=M_{2}(f,\III ). \notag
\end{equation*}

\subsection{Cohomologies of central extension of the Poisson
superalgebra} The following theorems are proved in the next sections.

\begin{theorem}\label{th4}$ $

Let $n_-=2k+1$. Let the bilinear mappings $\reve M_{2}^{1}$ and
$\reve M_{2}^{3}$ from $(\reve D^{n_-}_{n_+})^2$ to $\reve D^{n_
-}_{n_+}$ are defined by the relations \begin{eqnarray} \reve M_
{2}^1(z|f,g)&=&m_{2|1}(z|f,g), \ \reve{M}_{2}^1(f,\III )=0,\\ \reve
M_{2}^3(z|f,g)&=&m_{2|3}(z|f,g), \ \reve{M}_{2}^3(f,\III )=0,
\end{eqnarray} and the bilinear mapping $\reve L$ from $(\reve D^{1}_
{2})^2$ to $\reve D^{1}_{2}$ is defined by the relations
\begin{eqnarray} \reve L(z|f,g)&=&L_{2}^1(z|f,g) + \frac{20}{3}
\left(\int dz\xi f(z)\bar{g}- \int dz\xi g(z)\bar{f}\right)\III, \
\reve L (f,\III )=0, \end{eqnarray} where $m_{2|1}(z|f,g)$, $m_
{2|1}(z|f,g)$, and $L_2^1 (z|f,g)$ are defined by (\ref{2|1}),
(\ref{2|3}), and (\ref{L21}) respectively.

Let $V_2$ be the subspace of $C_2(\reve D^{n_-}_{n_+},\reve
D^{n_-}_{n_+})$ generated by the bilinear mappings $\reve M_
{2}^1$, $\reve M_{2}^{3}$, and $\delta_{k,0}\delta_{2,n_+} \reve
L$.

Then $dim V_2=2+\delta_{k,0}\delta_{2,n_+}$, and there is a
natural isomorphism $V_2\oplus (E^{2k+1}_{n_+}/D^{2k+1}_{n_
+})\simeq H^2_{\mathrm{ad}}$ taking $(\reve M_2,T)\in V_2\oplus
(E^{n_-}_{n_+}/D^{n_-}_{n_+})$ to the cohomology class
determined by the cocycle $$ \reve M_2(z|\reve f,\reve
g)-\{t(z),f(z)\}\bar{g}+
(-1)^{\varepsilon(f)\varepsilon(g)}\{t(z),g(z)\}\bar{f}, $$ where $t
\in E^{n_-}_{n_+}$ belongs to the equivalence class $T$.

\end{theorem}

\begin{theorem}\label{th5}$ $

Let $n_-=n_++4$. Let the bilinear mappings $\reve M_{2}^{3}$ and
${\reve Q}_{\ttt}$ from $(\reve D^{n_-}_{n_+})^2$ to $\reve D^{n_
-}_{n_+}$ are defined by the relations \begin{eqnarray} &&
\!\!\!\!\!\!\!\!\!\! \reve M_{2}^3(z|f,g)=\bar g \eE_z f-
(-1)^{\varepsilon(f)\varepsilon(g)} \bar f \eE_z g, \ \ \ \reve{M}_
{2}^3(\reve f,\III )=0, \\ && \!\!\!\!\!\!\!\!\!\! {\reve Q}_
{\ttt} (z|f, g)= \{\ttt (z),g(z)\}\bar{f}-(-1)^{\varepsilon
(f)\varepsilon (g)}\{\ttt (z),f(z)\}\bar{g} +\left( C_2^2(\ttt
,g)\bar{f}-(-1)^{\varepsilon (f)\varepsilon (g)}C_2^2(\ttt
,f)\bar{g} \right) \III \nonumber\\ && \ \ \ \ \ \ \ \ \ \ \ \reve
Q_\ttt(z|\reve f, \III)=0, \end{eqnarray} where $C_2^2$ is defined
by (\ref{c22}) and $\zeta\in E_{n_+}^{n_-}$.

Let $V_2$ be the subspace of $C_2(\reve D^{n_-}_{n_+},\reve
D^{n_-}_{n_+})$ generated by the bilinear mapping $\reve M_
{2}^{3}$.

Then $dim V_2=1$ and there is a natural isomorphism $V_2\oplus
(E^{2k+1}_{n_+}/D^{2k+1}_{n_+})\simeq H^2_{\mathrm{ad}}$ taking
$(\reve M_2,T)\in V_2\oplus (E^{n_-}_{n_+}/D^{n_-}_{n_+})$ to
the cohomology class determined by the cocycle $$ \reve M_2(z|\reve
f,\reve g)+\reve Q_\ttt (z|\reve f, \reve g) $$ where $\ttt \in
E^{n_-}_{n_+}$ belongs to the equivalence class $T$.

\end{theorem}

%%%%%%%%%%%%%%%%%%%%%%%%%%%%%%%%%%%%%%%

\subsection{Deformations of the Lie superalgebra}

Let $L$ be a topological Lie superalgebra over $\K$ with Lie
superbracket $\{\cdot,\cdot\}$, $\K[[\p]]$ be the ring of formal
power series in $\p$ over $\K$, and $L[[\p]]$ be the
$\K[[\p]]$-module of formal power series in $\p$ with coefficients in
$L$. We endow both $\K[[\p]]$ and $L[[\p]]$ by the direct-product
topology. The grading of $L$ naturally determines a grading of
$L[[\p]]$: an element $f=f_0+\p f_1+\ldots$ has a definite parity
$\varepsilon(f)$ if $\varepsilon(f)=\varepsilon(f_j)$ for all
$j=0,1,..$. Every $p$-linear separately continuous mapping from
$L^p$ to $L$ (in particular, the bracket $\{\cdot,\cdot\}$) is
uniquely extended by $\K[[\p]]$-linearity to a $p$-linear separately
continuous mapping over $\K[[\p]]$ from $L[[\p]]^p$ to $L[[\p]]$. A
(continuous) formal deformation of $L$ is by definition a
$\K[[\p]]$-bilinear separately continuous Lie superbracket
$C(\cdot,\cdot)$ on $L[[\p]]$ such that $C(f,g)=\{f,g\} \mod \p$ for
any $f,g\in L[[\p]]$. Obviously, every formal deformation $C$ is
expressible in the form \begin{equation} \label{1} C(f,g)=\{f,g\}+\p
C_1(f,g)+\pp{4}{2}C_2(f,g)+\ldots,\quad f,g\in L, \end{equation}
where $C_j$ are separately continuous skew-symmetric bilinear
mappings from $L\times L$ to $L$ (2-cochains with coefficients in the
adjoint representation of $L$). {}Formal deformations $C^1$ and
$C^2$ are called equivalent if there is a continuous
$\K[[\p]]$-linear operator $T: L[[\p]]\to L[[\p]]$ such that
$TC^1(f,g)=C^2(T f,Tg)$, $f,g\in L[[\p]]$.

Here as well as in \cite{deform}, we use more restricting definition

\begin{definition} {}Formal deformations $C^1$ and $C^2$ are called
similar if there are a continuous $\K[[\p]]$-linear operators $T$,
$T_1$: $L[[\p]]\to L[[\p]]$ such that $TC^1(f,g)=C^2(T f,T g)$,
$f,g\in L[[\p]]$ and $T f = f + \p T_1 f$. \end{definition}

\subsection{Deformations of central extension of the Poisson
superalgebra}

The following theorems are proved in subsequent sections:

\begin{theorem}\label{th6} Let $n_-=2k+1$. Let $ \mathcal{M}_
{\kappa }(z|f,g)=\frac{1}{\hbar \kappa }f(z)\sinh \left( \hbar \kappa
\frac{\overleftarrow{\partial }}{\partial z^{A}}\omega ^{AB}\frac{
\partial }{\partial z^{B}}\right) g(z). $ {}For given $\kappa$, let
$\zeta (z),w(z)\in E^{2k+1}_{n_+}[[\p]]$ satisfy the following
conditions \begin{eqnarray} i) &&\varepsilon (\zeta )=1, \ \
\varepsilon (w)=0; \label{i}\\ ii) &&\mathcal{M}_{\kappa
}(z|\zeta ,\zeta )+w(z)\in D^{2k+1}_{n_+}[[\p]]. \label{ii}
\end{eqnarray} Let $ \reve{N}_{2|\kappa ,\zeta
,w}(\reve{f},\reve{g})=N_{2|\kappa ,\zeta ,w}(z|
\reve{f},\reve{g})+n_{2|\kappa ,\zeta ,w}(\reve{f},\reve{g})\III , $
where \begin{eqnarray*} && N_{2|\kappa ,\zeta ,w}(z|f,g)=
\mathcal{M}_{\kappa }(z|f+\p\zeta \bar{f},g+\p\zeta \bar{g})+
\hbar^4 w(z)\bar{f}\bar{g},\\ && N_{2|\kappa ,\zeta ,w}(z|f,\III )=
\mathcal{M}_{\kappa }(z|w,f+\p \zeta \bar{f}),\\ && n_{2|\kappa
,\zeta ,w}(f,g)=\bar{f}\bar{g},\;\\ && n_{2|\kappa ,\zeta ,w}(f,\III
)=0,\;\\ && \reve{N}_{2|\kappa ,\zeta ,w}(z|\III ,\III )=0.
\end{eqnarray*} Then \begin{enumerate} \item every continuous formal
deformation of the superalgebra $\reve D_{n_+}^{2k+1}$ is similar
to the superbracket $$ [ \reve{f},\reve{g}]_{\kappa,\zeta,w }
 =\reve{N}_{2|\kappa, \zeta,w}(\reve{f},\reve{g}) $$ with some
$\kappa$ and $\zeta (z),w(z)\in E^{2k+1}_{n_+}[[\p]]$ satisfying
 (\ref{i})-(\ref{ii}) ; \item the superbracket $[ \reve{f},\reve{g}]_
{\kappa,\zeta_1,w_1 }$ is similar to the superbracket $[
\reve{f},\reve{g}]_{\kappa,\zeta,w }$, if $\zeta_1(z)-\zeta(z)\in
D_{n_+}^{n_-}[[\p]]$ and $w_1(z)-w(z)\in D_{n_+}^{n_-}[[\p]]$.
\end{enumerate}

\end{theorem}

\begin{theorem}\label{th7} Let $n_-=n_++4$. Let $\zeta(z)\in E_{n_
+}^{n_++4}[[\p]]$, $c_3\in \K[[\p]]$, $C(f,g)=\int
dz\left(f(z){}\eE_z g(z)- (-1)^{\varepsilon(f)\varepsilon(g)}
g(z){}\eE_z f(z)\right)$ and \begin{eqnarray*} &&\reve{S}_{2|\zeta
,c_{3}}(\reve{f},\reve{g}) =S_{2|\zeta ,c_{3}}(z|
\reve{f},\reve{g})+s_{2|\zeta ,c_{3}}(\reve{f},\reve{g})\III , \\
&&S_{2|\zeta ,c_{3}}(z|f,g) =\left\{ f(z),g(z)\right\} +c_
{3}\left( \bar{f}\mathcal{E}_{z}g(z)-(-1)^{\varepsilon
(f)\varepsilon (g)} \bar{g}\mathcal{E}_{z}f(z) \right) + \\
&&\qquad \qquad \qquad + \p\left(\{\zeta
(z),g(z)\}\bar{f}-(-1)^{\varepsilon(f)\varepsilon (g)} \{\zeta
(z),f(z)\}\bar{g}\right) , \\ &&s_{2|\zeta,c_
{3}}(f,g)=C(f,g)+ C(\p\zeta ,g)\bar{f}-(-1)^{\varepsilon
(f)\varepsilon (g)}C(\p\zeta ,f)\bar{g} , \\ &&S_{2|\zeta ,c_
{3}}(\reve{f},\III ) =s_{2|\zeta ,c_{3}}(\reve{f},\III )=0.
\end{eqnarray*}

Then \begin{enumerate} \item every continuous formal deformation of
the superalgebra $\reve D_{n_+}^{n_++4}$ is similar to the
superbracket \begin{equation*} [ \reve{f},\reve{g}]_{\zeta ,c_{3}
}=\reve{S}_{2|\zeta ,c_{3}}(\reve{f},\reve{g}) \end{equation*} with
some $c_3$ and $\zeta (z)\in E^{2k+1}_{n_+}[[\p]]$ ;

\item the superbracket $[ \reve{f},\reve{g}]_{\zeta_1 ,c_{3} }$ is
similar to the superbracket $[ \reve{f},\reve{g}]_{\zeta ,c_{3} }$,
if $\zeta_1(z)-\zeta(z)\in D_{n_+}^{n_-}[[\p]]$.

\end{enumerate} \end{theorem}

%%%%%%%%%%%%%%%%%%%%%%%%%%%%%%%%%%%%

\section{Superalgebra $\reve D_{n_+}^{2k+1}$} \subsection{Second
adjoint cohomology}\label{sectionodd}

In the superalgebra under consideration we have
$C(f,g)=\bar{f}\bar{g}$ and $ [f,g]=\{f,g\}+\bar{f}\bar{g}\III $.

Consider the cohomology equation $ \reve{d}_
{2}^{\mathrm{ad}}\reve{M}_{2}(\reve{f},\reve{g},\reve{h})=0. $ It
follows from (\ref{1.3}) that $ d_{1}^{\mathrm{ad}}\tilde{M}_
{1}(z|f,g)=0 $ and so, due to Theorem \ref{th0} \be M_{2}(z|f,\III )
=t^{0}\mathcal{E}_{z}f(z)+\left\{ t^{\prime }(z),f(z)\right\} ,\ \
t^{\prime }(z)\in E^{2k+1}_{n_+},\ \ \varepsilon (t^{0})
=\varepsilon (t^{\prime }(z))=\varepsilon _{M_{2}}, \ee

It follows from (\ref{1.3}) also that $ m_{2}(\left\{ f,g\right\}
,\III ) =(2+n_{+}-n_{-})t^{0}\bar{f}\bar{g}$ which implies $t^0=0$
and $m_{2}(\left\{ f,g\right\} ,\III )=0$. In its turn, this gives
$ m_{2}(f, \III) =m_{2}\bar{f},\;m_
{2}=\mathrm{const},\;\varepsilon (m_{2})=\varepsilon _{M_{2}}+1.
$

So, we have proved \begin{proposition} \label{prop6}
\begin{eqnarray*} &&M_{2}(z|f,\III )=\left\{ t^{\prime
}(z),f(z)\right\} ,\;m_{2}(f,\III )=m_{2}\bar{f}, \\ &&t^{\prime
}(z)\in E_{n_+}^{2k+1},\;\varepsilon (t^{\prime }(z))=\varepsilon _
{M_{2}},\;\varepsilon (m_{2})=\varepsilon _{M_{2}}+1.
\end{eqnarray*} \end{proposition}

{}Further, it follows from (\ref{1.2}) that $ d_{2}^{\mathrm{ad}}M_
{2}^{\prime }(z|f,g,h)=0, $ where $ M_{2}^{\prime }(z|f,g)=M_
{2}(z|f,g)-\bar{f}\bar{g}t^{\prime }(z) $.

So, the proposition follows from Theorem \ref{th2} and the condition
$M_{2}(z|f,g)\in D^{2k+1}_{n_+}$ \begin{proposition}\label{prop7}
\begin{eqnarray*} M_{2}(z|f,g)&=&c_{1}m_{2|1}(z|f,g)+ t_{D}(z)m_
{2|2}(z|f,g)+c_{3}m_{2|3}(z|f,g)+m_{2|\zeta }(z|f,g)\- +\\ &+&
c^k_{5}m_{2|5}(z|f,g)+d_{1}^{\mathrm{ad}}\varphi_{D}(z|f,g),
\end{eqnarray*} where \begin{eqnarray*} &&m_{2|1}(z|f,g)=f(z)\left(
\overleftarrow{\partial }_{A}\omega ^{AB}\partial _{B}\right)
^{3}g(z),\;\bar{m}_{2|1}(|f,g)=0, \ \ \varepsilon _{m_{2|1}}=0,\\
&&m_{2|2}(z|f,g)=\bar{f}\bar{g}, \\ &&m_
{2|3}(z|f,g)=\bar{f}\mathcal{E}_{z}g(z)-(-1)^{\varepsilon
(f)\varepsilon (g)} \bar{g}\mathcal{E}_{z}f(z),\;\bar{m}_
{2|3}(|f,g)=0, \ \ \varepsilon _{m_{2|3}}=n_-, \\ &&t_
{D}(z)=t^{\prime }(z)+c_{2}\in D^{2k+1}_{n_+},\;\varepsilon (t_
{D}(z))=\varepsilon _{M_{2}}, \\ &&m_{2|\zeta }(z|f,g)=\{\zeta
(z),g(z)\}\bar{f}-(-1)^{\varepsilon (f)\varepsilon (g)}\{\zeta
(z),f(z)\}\bar{g},\;\bar{m}_{2|\zeta }(|f,g)=0,\; \\ &&\ \ \ \ \ \ \
\ \zeta (z)\in E^{2k+1}_{n_+}/D^{2k+1}_{n_+} ,\;\varepsilon
(\zeta (z))=\varepsilon _{M_{2}}+n_-, \\ &&m_{2|5}(z|f,g)=\delta
_{n_{+},2}L_2^{2k+1},\ \ \varepsilon _{m_{2|5}}=1, \ \ L_
2^{2k+1}=0 \mbox { for } k>1, \\ &&\varphi_{D}(z|f)\in C_
1(D^{2k+1}_{n_+},D^{2k+1}_{n_+}),\ \ \varepsilon _{\varphi_
{D}}=\varepsilon _{M_{2}}, \end{eqnarray*} \begin{eqnarray*}
&&\bar{M}_{2}(|f,g)=\bar{t}_{D}\bar{f}\bar{g}+c^k_{5}a_{k}\omega
(f,g)- \bar{\zeta}(|\left\{ f,g\right\} ), \\ &&\omega (f,g)=\delta _
{n_{+},2}\int dxd\eta d\xi f(x,\xi )g(x,\eta ),\; a_{k}= \left\{
\begin{array}{c} \frac{20}{3},\;k=0, \\ 6,\;k=1, \\ 0,\;k\geq 2,
\end{array} \right. \end{eqnarray*} and the bilinear forms $L_2^i$
are defined by (\ref{L21}). \end{proposition}

To find $m_{2}(f,g)$ and the relations between the parameters in
Proposition ~\ref{prop7} , consider Eq. (\ref{1.2}). It gives
\begin{eqnarray} \label{eeq} d_{2}^{\mathrm{tr}}m_{2}^{\prime
}(f,g,h) &=&3(m_{2}-\bar{t}_{D}) \bar{f}\bar{g}\bar{h} -c^k_{5}a_
{n_{-}}[\omega (f,g)\bar{h}+\omega (f,h)\bar{g}+\omega (g,h)
\bar{f}], \end{eqnarray} where \begin{eqnarray*} m_{2}^{\prime
}(f,g) &=&m_{2}(f,g)-[\bar{\varphi}_{D}(|f)\bar{g}
-(-1)^{\varepsilon (f)\varepsilon (g)}\bar{\varphi}_{D}(|g)\bar{f}
]. \end{eqnarray*} Evidently, there exist such functions $f,g,h\in
D_{n_+}^{2k+1}$ that $\bar{f}\bar{g}\bar{h}\ne 0$,
$\{f,g\}=\{g,h\}=\{h,f\}=0$ and $\omega (f,g)=\omega (g,h)=\omega
(h,f)=0$. Indeed, let $$\supp (f)\bigcap \supp (g)=\supp (g)\bigcap
\supp (h)= \supp (h)\bigcap \supp (f)=\varnothing.$$ So, Eq.
(\ref{eeq}) implies \begin{proposition}\label{prop8} $ m_{2}
=\bar{t}_{D}, $ \end{proposition} and as a consequence
\begin{equation} d_{2}^{\mathrm{tr}}m_{2}^{\prime }(f,g,h)=-c^k_
{5}a_{k}[\omega (f,g)\bar{h} +\omega (f,h)\bar{g}+\omega
(g,h)\bar{f}] \label{2.2.1} \end{equation}

\begin{proposition}\label{prop9} $c_5^k=0$ for $k\ge 1$, i.e. $c_
5^k=c_5 \delta_{0,k}, $ where $c_5\in \K$ \end{proposition}
\begin{proof} Consider Eq. (\ref{2.2.1}) for the functions
$f=g=h=\varphi (x)\delta(\xi)$ \end{proof}

\begin{proposition}\label{prop11} Let $k=0$. Then \be \omega
(f,g)\bar{h}+\omega (f,h)\bar{g}+\omega (g,h)\bar{f}= \frac{3}{20}d_
{2}^{\mathrm{tr}}\chi _{2}(f,g,h), \ee where \be \chi _
{2}(f,g)=\frac{20}{3} \delta _{n_{+},2} \left(\int dz\xi
f(z)\bar{g}- \int dz\xi g(z)\bar{f}\right). \ee \end{proposition}
\begin{proof} \begin{eqnarray*} &&\omega (f,g)\bar{h}+\omega
(f,h)\bar{g}+\omega (g,h)\bar{f}= \delta_{n_{+},2}\Big(\int dxf_
{n_{-}}(x)g_{n_{-}}(x)\int dyh_{n_{-}}(y)+ \\ &&\,+\int dxf_{n_
{-}}(x)h_{n_{-}}(x)\int dyg_{n_{-}}(y)+ \int dx g_{n_{-}}(x)h_
{n_{-}}(x)\int dyf_{n_{-}}(y)\Big)= \\ &&\,=\delta
_{n_{+},2}\left(\int dz\xi \{f(z),g(z)\}\bar{h}+ \mathrm{cycle}
(f,g,h)\right) \end{eqnarray*}

\end{proof}

Thus, we obtain

\begin{eqnarray*} &&M_{2}(z|f,g)=c_{1}m_{2|1}(z|f,g)+c_{3}m_
{2|3}(z|f,g)+ m_{2|\zeta}(z|f,g)+c_{5}\delta_{k,0}\delta_{2,n_
+}L_2^1(z|f,g)+ \\ &&+t_D(z)m_{2|2}(z|f,g)+d_
{1}^{\mathrm{ad}}\varphi (z|f,g), \\ &&m_{2}(f,g)=\bar{\varphi}_
D(|f)\bar{g}-(-1)^{\varepsilon (f)\varepsilon (g)} \bar{\varphi}_
D(|g)\bar{f}+c_{5}\delta_{k,0}\chi _{2}(f,g)+ b_
{2}\bar{f}\bar{g}+\mu_{1}(\left\{ f,g\right\} ), \\ &&M_
{2}(z|f,\III )=\left\{ t_D(z),f(z)\right\} ,\;m_{2}(f,\III
)=\bar{t} \bar{f}. \end{eqnarray*} where $\mu_1\in C_1 (D_{n_
+}^{2k+1},\K)$, $t_D \in D_{n_+}^{2k+1}$, $\varphi_D \in C_1 (D_
{n_+}^{2k+1}, D_{n_+}^{2k+1})$.

Let $\reve{M}_{1D}(\reve{f})= \varphi_D (z|f)-\mu_{1}(f) \III$,
$\reve{M}_{1D} (\III )=-t_{D}(z)-b_{2}\III $. Then \be \reve{M}_
{2}(\reve{f},\reve{g})=\reve{M}_{2|1}(\reve{f},\reve{g})+ \reve{d}_
{1}^{\mathrm{ad}}\reve{M}_{1D}(\reve{f},\reve{g}) \ee where
\begin{eqnarray*} &&\reve{M}_{2|1}(f,g)=c_{1}m_{2|1}(z|f,g)+c_
{3}m_{2|3}(z|f,g)+ m_{2|\zeta } (z|f,g)+c_{5}\delta_{k,0}\delta_
{2,n_+}L_2^1(z|f,g)+ \\ &&+c_{5}\delta_{k,0}\chi _{2}(f,g)\III ,
\\ &&\reve{M}_{2|1}(f,\III )=0 \end{eqnarray*}

{}Finally, we have, that up to coboundary, the cocycle $\reve{M}_
{2}(\reve{f},\reve{g})$ has the form \begin{eqnarray} &&M_
{2}(z|f,g)=c_{1}m_{2|1}(z|f,g)+c_{3}m_{2|3}(z|f,g)+ m_
{2|\zeta}(z|f,g)+c_{5}\delta_{k,0}\delta_{2,n_+}L_2^1(z|f,g), \\
&&m_{2}(f,g)=\delta_{k,0}c_{5}\chi_{2}(f,g), \\ &&\reve{M}_
{2}(f,\III )=0. \end{eqnarray}

\subsubsection{Independence and non-triviality}

Suppose that \begin{equation*} \reve{M}_
{2}(\reve{f},\reve{g})=\reve{d}_{1}^{\mathrm{ad}}\reve{M}_{1}(
\reve{f},\reve{g}). \end{equation*} This relation yields
\begin{equation} c_{1}m_{2|1}(z|f,g)+c_{3}m_{2|3}(z|f,g)+c_
{5}\delta_{k,0}\delta_{2,n_+}L_2^1(z|f,g) =d_{1}^{\mathrm{ad
}}M_{1}(z|f,g)-\bar{f}\bar{g}M_{1}(z|\III ). \label{15}
\end{equation} Let \begin{equation*} z\bigcap \left[
\mathrm{supp}(f)\bigcup \mathrm{supp}(g)\right] = \mathrm{supp
}(f)\bigcap \mathrm{supp}(g)=\varnothing . \end{equation*} It
follows from (\ref{15}) \begin{equation*} \bar{f}\bar{g}M_{1}(z|\III
)=0\;\Longrightarrow \;M_{1}(z|\III )=0\;\Longrightarrow
\end{equation*} \begin{equation*} c_{1}m_{2|1}(z|f,g)+c_{3}m_
{2|3}(z|f,g)+c_{5}\delta_{k,0}\delta_{2,n_+}L_2^1(z|f,g)= d_
{1}^{\mathrm{ad}}M_{1}(z|f,g). \end{equation*} This equation has
solution for $c_{1}=c_{3}=c_{5}=0$ only$.$

\subsection{Deformations of Lie superalgebra $\reve D_{n_
+}^{2k+1}$}

In this section, we find the general form of the deformation of Lie
superalgebra $\reve D_{n_+}^{2k+1}$, $ [f,g]_{\ast }$,
\begin{eqnarray*} &[\reve{f},\reve{g}]_{\ast }=&M_{2}^{\ast
}(z|\reve{f},\reve{g} )+m_{2}^{\ast }(\reve{f},\reve{g})\III = \sum_
{l=0}^\infty \hbar ^{2l} \reve{M}_{2l}(\reve{f},\reve{g}), \\ &&M_
{2}^{\ast }(z|\reve{f},\reve{g})\in D^{2k+1}_{n_+}[[\p]],\; m_
{2}^{\ast }(\reve{f},\reve{ g})\in \K[[\p]],\;\varepsilon _{M_
{2}^{\ast }}=\varepsilon _{m_{2}^{\ast }}=0, \end{eqnarray*}
satisfying the Jacoby identity. \begin{equation} (-1)^{\varepsilon
(\reve{f})\varepsilon (\reve{h})}[[\reve{f},\reve{g} ]_{\ast
},\reve{h}]_{\ast }+(-1)^{\varepsilon (\reve{g})\varepsilon (
\reve{h})}[[\reve{h},\reve{f}]_{\ast },\reve{g}]_{\ast
}+(-1)^{\varepsilon (\reve{f})\varepsilon
(\reve{h})}[[\reve{g},\reve{h} ]_{\ast },\reve{f}]_{\ast }=0.
\label{2.3.1} \end{equation}

We have $ M_{20}(f,g)=\left\{ f(z),g(z)\right\} ,\;m_
{20}(f,g)=\bar{f}\bar{g},\; M_{20}(\reve{f},\III )=m_
{20}(\reve{f},\III )=0. $

{}For any $\kappa \in \K$, the Moyal-type superbracket
\begin{equation*} \mathcal{M}_{\kappa }(z|f,g)=\frac{1}{\hbar \kappa
}f(z)\sinh \left( \hbar \kappa \frac{\overleftarrow{\partial
}}{\partial z^{A}}\omega ^{AB}\frac{ \partial }{\partial
z^{B}}\right) g(z) \end{equation*} is antisymmetric and satisfies the
Jacobi identity. \begin{definition} {}For $\zeta \in E^{2k+1}_{n_
+} $, $\kappa \in \K$, we set \begin{equation} \mathcal{N}_{\kappa
,\zeta }(z|f,g)=\mathcal{M}_{\kappa }(z|f+\zeta \bar{f} ,g+\zeta
\bar{g}). \notag \end{equation} \end{definition} It is shown in
\cite{deform}, that if $\zeta_1$ and $\zeta_2$ belong to the same
equivalence class of $E^{2k+1}_{n_+}/D^{2k+1}_{n_+}$ and
$\mathcal{N}_{\kappa ,\zeta_1 }(z|f,g) \in D^{2k+1}_{n_+}[[\p]] $
for all $f,g\in D^{2k+1}_{n_+}[[\p]]$ then the bilinear forms
$\mathcal{N}_{\kappa ,\zeta_1 }(z|f,g)$ and $\mathcal{N}_{\kappa
,\zeta_2 }(z|f,g)$ are equivalent under some similarity
transformation $T$, mapping $D^{2k+1}_{n_+}[[\p]]$ to $D^{2k+1}_
{n_+}[[\p]]$, \be \mathcal{N}_{\kappa ,\zeta_1 }(z| T f,T g)= T
\mathcal{N}_{\kappa ,\zeta_2 }(z|f,g). \ee

\subsubsection{$\hbar ^{2}$-order}

In $\hbar ^{2}$-order, Eq. (\ref{2.3.1}) gives an equation
\begin{equation*} \reve{d}_{2}^{\mathrm{ad}}\reve{M}_
{21}(\reve{f},\reve{g},\reve{h} )=0,\;\varepsilon _{\reve{M}_
{21}}=0, \end{equation*} general solution of which is \footnote{ We
should set $c_3 = c_{5} = 0$ because $\varepsilon _{m_
{2|3}}=\varepsilon _{L_2^1}=1$.} \begin{eqnarray*} &&\reve{M}_
{21}(\reve{f},\reve{g})=\reve{M}_{2|11}(\reve{f},\reve{g})+
\reve{d}_{1}^{\mathrm{ad}}\reve{M}_{1D1}(\reve{f},\reve{g}), \\
&&M_{2|11}(z|f,g)=\frac{1}{6}\kappa _{1}^{2}m_{2|1}(z|f,g)+m_
{2|\zeta _{1}}(z|f,g), \\ &&\varepsilon (\zeta _{1}(z))=1,\;\zeta _
{1}(z)\in E^{2k+1}_{n_+}/D^{2k+1}_{n_+} , \\ &&m_
{2|11}(f,g)=\reve{M}_{2|11}(f,\III )=0, \\ &&\reve{M}_
{1D1}(f)=\zeta _{D1}(z|f)+m_{11}(f)\III ,\;\reve{M} _{1D1}(\III
)=t_{D1}(z)+b_{1}\III . \end{eqnarray*} Performing the similarity
transformation $[\reve{f},\reve{g}]_{\ast }\rightarrow [
\reve{f},\reve{g}]_{\ast T}$ with\footnote{ Note that
$\reve{T}=1+\hbar ^{2}\reve{M}_{1}+O(\hbar ^{4})$ is a similarity
transformation if $M_{1}(z|f),M_{1}(z|\III ) \in D^{2k+1}_{n_
+}[[\p]]$ for any $f\in D^{2k+1}_{n_+}[[\p]]$. }
$\reve{T}(\reve{f} )=\reve{f}-\hbar ^{2}\reve{M}_
{1D1}(\reve{f})+O(\hbar ^{4})$, we can rewrite $ [\reve{f},\reve{g}]_
{\ast }$ in the form \begin{equation*} [ \reve{f},\reve{g}]_{\ast
}=\reve{N}_{\kappa _{[ 1]},\zeta _{[ 1]}}(\reve{f},\reve{g})+\hbar
^{4}\reve{M}_{22}(\reve{f}, \reve{g})+O(\hbar ^{6}), \end{equation*}
where \begin{eqnarray*} &&\reve{N}_{\kappa _{[ 1]},\zeta _{[
1]}}(f,g)=\mathcal{N} _{\kappa _{[ 1]},\zeta _{[
1]}}(z|f,g)-\mathcal{M}_{\kappa [1] }(z|\zeta _{[ 1]},\zeta _{[
1]})_{2}\bar{f}\bar{g}+\bar{f}\bar{g} \III , \\ &&\mathcal{\bar{N}}_
{\kappa _{[ 1]},\zeta _{[ 1]}}(|f,g)- \mathcal{\bar{M}}_{\kappa
[1]}(|\zeta _{[ 1]},\zeta _{[ 1]})_{2} \bar{f}\bar{g}=O(\hbar
^{6}), \\ &&\reve{N}_{\kappa _{[ 1]},\zeta _{[ 1]}}(f,\III
)=\reve{N} _{\kappa _{[ 1]},\zeta _{[ 1]}}(\III ,\III )=0,
\end{eqnarray*} \begin{equation*} \kappa _{[ n]}=\sum_
{k=1}^{n}\hbar ^{2(k-1)}\kappa _{k},\;\zeta _{[ n]}=\sum_
{k=1}^{n}\hbar ^{2k}\zeta _{k}, \end{equation*} and $\hbar
^{2(n+1)}\mathcal{M}_{\kappa [n]}(z|\zeta _{[ n]},\zeta _{[ n]})_
{n+1}$ is the $\hbar ^{2(n+1)}$-order term of expansion of
$\mathcal{M} _{\kappa [n]}(z|\zeta _{[ n]},\zeta _{[ n]})$ in
$\hbar ^{2}$ -series.

The condition $[\reve{f},\reve{g}]_{\ast }\in \reve D_{n_
+}^{2k+1}$ implies that $\reve{M}_{22}(\reve{f},\reve{g})\in \reve
D_{n_+}^{2k+1}$.

\subsubsection{$\hbar ^{4}$-order}

The Jacobi identity (\ref{2.3.1}) gives the following equations in
$\hbar ^{4}$-order \begin{equation} d_{1}^{\mathrm{ad}}\tilde{M}_
{12}(z|f,g)=0,\;\tilde{M} _{12}(z|f)=M_{22}(z|f,\III ),
\label{2.3.2.1} \end{equation} \begin{equation} \bar{f}\bar{M}_
{22}(|g,\III )-(-1)^{\varepsilon (f)\varepsilon (g)}\bar{g} \bar{M}_
{22}(|f,\III )+m_{22}(\left\{ f,g\right\} ,\III )=0.
\label{2.3.2.2} \end{equation} \begin{eqnarray} &&\!\!\!\!\!\! d_
{2}^{\mathrm{ad}}M_{22}^{\prime }(z|f,g,h)+[\bar{f}\bar{g} M_
{22}(z|h,\III )-(-1)^{\varepsilon (f)\varepsilon (g)}\bar{f}\bar{h}
M_{22}(z|g,\III )+\bar{g}\bar{h}M_{22}(z|f,\III )]=0,
\label{2.3.2.3} \\ && \!\!\!\!\!\! M_{22}^{\prime }(z|f,g)=M_
{22}(z|f,g)-\mathcal{M}_{\kappa [1]}(z|\zeta _{[ 1]},\zeta _{[
1]})_{2}\bar{f}\bar{g}, \notag \end{eqnarray} \begin{eqnarray} &&d_
{2}^{\mathrm{tr}}m_{22}(f,g,h)=(-1)^{\varepsilon (f)\varepsilon
(h)}[(-1)^{\varepsilon (f)\varepsilon (h)}\bar{f}\bar{M}_{22}(|g,h)
\notag \\ &&+(-1)^{\varepsilon (f)\varepsilon (h)}\bar{f}\bar{g}m_
{22}(h,\III )+ \mathrm{cycle}(f,g,h)]. \label{2.3.2.4}
\end{eqnarray} The general solution of Eq. (\ref{2.3.2.1}) has the
form \begin{eqnarray*} &&M_{22}(z|f,\III )=t_{2}^{0}\mathcal{E}_
{z}f(z)+\left\{ t_{2}(z),f(z)\right\} ,\; \\ &&\varepsilon (t_
{2}^{0})=\varepsilon (t_{2}(z))=0,\;t_{2}(z)\in E^{2k+1}_{n_+}.
\end{eqnarray*} {}Further, Eq. (\ref{2.3.2.2}) gives
\begin{equation*} t_{2}^{0}=0,\;m_{22}(f,\III )=m_{22}\bar{f},\;m_
{22}=\mathrm{const}, \end{equation*} such that \begin{equation*} M_
{22}(z|f,\III )=\left\{ t_{2}(z),f(z)\right\} ,\;m_{22}(f,\III )=m_
{22} \bar{f}. \end{equation*} The condition $\varepsilon _{m_
{22}}=0$ gives $m_{22}=0$ and \begin{equation*} m_{22}(f,\III )=0.
\end{equation*} {}Further, we find from Eq. (\ref{2.3.2.3})
\begin{eqnarray*} &&M_{22}(f,g)=\frac{\kappa _{1}\kappa _{2}}{3}m_
{2|1}(z|f,g)+m_{2|\zeta _{2}}(z|f,g)+ \\ &&+\left( \mathcal{M}_
{\kappa [1]}(z|\zeta _{[ 1]},\zeta _{[ 1]})_{2}+t_{2}(z)+c_
{22}\right) \bar{f}\bar{g}+d_{1}^{\mathrm{ad}}\zeta _{D2}(z|f,g),
\\ &&\mathcal{M}_{\kappa [1]}(z|\zeta _{[ 1]},\zeta _{[ 1]})_
{2}+t_{2}(z)+c_{22}\in D^{2k+1}_{n_+}, \;\zeta _{D2}(z|f)\in
D^{2k+1}_{n_+},\;\zeta _{2}(z)\in E^{2k+1}_{n_+}/D^{2k+1}_{n_
+} . \end{eqnarray*} {}Finally, we obtain from Eq. (\ref{2.3.2.4})
\begin{equation*} m_{22}(f,g)=\bar{\zeta}_
{D2}(|f)\bar{g}-(-1)^{\varepsilon (f)\varepsilon (g)}\bar{\zeta}_
{D2}(|g)\bar{f}-b_{22}\bar{f}\bar{g}-\mu _{12}(\left\{ f,g\right\}
). \end{equation*} Introduce notation \begin{equation*} t_{2}(z)+c_
{22}=t_{D2}(z)+w_{2}(z),\;t_{D2}(z)\in D^{2k+1}_{n_+},\;w_
{2}(z)\in E^{2k+1}_{n_+}/D^{2k+1}_{n_+} . \end{equation*} Then,
we can write \begin{equation*} \reve{M}_
{22}(\reve{f},\reve{g})=\reve{M}_{22|\mathrm{co}}(\reve{f},
\reve{g})+\reve{d}_{1}^{\mathrm{ad}}\reve{M}_
{1D2}(\reve{f},\reve{g}), \end{equation*} \begin{eqnarray*} &&M_
{22|\mathrm{co}}(f,g)=\frac{\kappa _{1}\kappa _{2}}{3} m_
{2|1}(z|f,g)+m_{2|\zeta _{2}}(z|f,g)+ \\ &&+\left( \mathcal{M}_
{\kappa [1]}(z|\zeta _{[ 1]},\zeta _{[ 1]})_{2}+w_{2}(z)\right)
\bar{f}\bar{g},\;\left( \mathcal{M}_{\kappa [1] }(z|\zeta _{[
1]},\zeta _{[ 1]})_{2}+w_{2}(z)\right) \in D^{2k+1}_{n_+},
\end{eqnarray*} \begin{equation*} M_{22|\mathrm{co}}(z|f,\III
)=\left\{ w_{2}(z),f(z)\right\} ,\;m_{22|\mathrm{co}}(f,g)=m_
{22|\mathrm{co}}(f,\III )=0. \end{equation*} \begin{equation*} M_
{1D2}(z|f)=\zeta _{D2}(z|f),\;M_{12}(z|\III )=t_{D2}(z),\;m_
{12}(f)=\mu _{12}(f),\;m_{12}(\III )=b_{22}. \end{equation*}

\subsubsection{Higher orders} Introduce a form $\reve{N}_{2|\kappa
,\zeta ,w}(\reve{f},\reve{g})$, \begin{equation*} \reve{N}_{2|\kappa
,\zeta ,w}(\reve{f},\reve{g})=N_{2|\kappa ,\zeta ,w}(z|
\reve{f},\reve{g})+n_{2|\kappa ,\zeta ,w}(\reve{f},\reve{g})\III ,
\end{equation*} where \begin{eqnarray*} &&N_{2|\kappa ,\zeta
,w}(z|f,g)=\mathcal{M}_{\kappa }(z|f+\zeta \bar{f} ,g+\zeta
\bar{g})+w(z)\bar{f}\bar{g}, \\ &&N_{2|\kappa ,\zeta ,w}(z|f,\III )=
\mathcal{M}_{\kappa }(z|w,f+\zeta \bar{f}),\\ && n_{2|\kappa ,\zeta
,w}(f,g)=\bar{f}\bar{g},\;n_{2|\kappa ,\zeta ,w}(f,\III
)=0,\;\reve{N}_{2|\kappa ,\zeta ,w}(z|\III ,\III )=0, \\ &&
\varepsilon (\zeta )=1,\;\varepsilon (w)=0, \\ &&
\zeta (z),w(z)\in E^{2k+1}_{n_+}[[\p]]/D^{2k+1}_{n_+}[[\p]],
\\ && \psi=\mathcal{M}_{\kappa }(z|\zeta ,\zeta )+w(z)\in D^{2k+1}_
{n_+}[[\p]], \end{eqnarray*} Note that it follows from the Jacoby
identity for the form $\mathcal{M}_{\kappa }(z|f,g)$ that
\begin{equation*} \mathcal{M}_{\kappa }(z|\mathcal{M}_{\kappa
}(|\zeta ,\zeta ),\zeta )\equiv 0, \end{equation*} such that
\begin{eqnarray*} \mathcal{M}_{z\kappa }(|w,\zeta ) &=&\mathcal{M}_
{\kappa }(z|\mathcal{M} _{\kappa }(|\zeta ,\zeta )+w,\zeta
)=\mathcal{M}_{\kappa }(z|\psi ,\zeta )\in D_{n_+}^{2k+1}[[\p]],
\\ \mathcal{\bar{M}}_{\kappa }(z|w,\zeta ) &=&\mathcal{\bar{M}}_
{\kappa }(z|\psi ,\zeta )=0. \end{eqnarray*} Analogously, the
condition \begin{equation*} \mathcal{M}_{\kappa _{\lbrack
n]}}(z|\zeta _{\lbrack n]},\zeta _{\lbrack n]})_{[n+1]}+w(z)_
{[n+1]}\in D_{n_+}^{2k+1}[[\p]],\; w(z)_{[n]}=\sum_
{k=2}^{n}\hbar^{2k}w_{k}(z), \end{equation*} implies
\begin{equation*} \mathcal{M}_{\kappa _{\lbrack n+1]}}(z|w(z)_
{[n+1]},\zeta _{\lbrack n+1]})_{[n+2]}\in D_{n_
+}^{2k+1}[[\p]],\;\mathcal{\bar{M}}_{\kappa _{\lbrack
n+1]}}(z|w(z)_{[n+1]},\zeta _{\lbrack n+1]})_{[n+2]}=0.
\end{equation*}

Straightforward calculations give that the form $\reve{N}_{2|\kappa
,\zeta ,w}(\reve{f},\reve{g})$ satisfies the Jacoby identity
\begin{eqnarray*} &&(-1)^{\varepsilon (\reve{f})\varepsilon
(\reve{h})}\reve{N}_{2|\kappa ,\zeta ,w}(\reve{f},,\reve{N}_
{2|\kappa ,\zeta ,w}(\reve{g},\reve{h} ))+(-1)^{\varepsilon
(\reve{h})\varepsilon (\reve{g})}\reve{N}_{2|\kappa ,\zeta
,w}(\reve{h},\reve{N}_{2|\kappa ,\zeta ,w}(\reve{f},\reve{g}))+ \\
&&\,+(-1)^{\varepsilon (\reve{g})\varepsilon (\reve{f})}\reve{N} _
{2|\kappa ,\zeta ,w}(\reve{g},\reve{N}_{2|\kappa ,\zeta
,w}(\reve{h}, \reve{f}))=0. \end{eqnarray*}

Performing the similarity transformation $[\reve{f},\reve{g}]_{\ast
}\rightarrow [ \reve{f},\reve{g}]_{\ast T}$ with
$\reve{T}(\reve{f})=\reve{f}-\hbar ^{4}\reve{M}_{1D2}(\reve{f}
)+O(\hbar ^{6})$, one can rewrite $[\reve{f},\reve{g}]_{\ast }$ in
the form \begin{eqnarray*} [ \reve{f},\reve{g}]_{\ast } &=&\reve{N}_
{2|\kappa _{[ 2]},\zeta _{[ 2]},w_
{[2]}}(\reve{f},\reve{g})-\mathcal{L}_{23}(z|
\reve{f},\reve{g})+\hbar ^{6}\reve{M}_
{23}(\reve{f},\reve{g})+O(\hbar ^{8}),\; \\ \mathcal{L}_{23}(z|f,g)
&=&\mathcal{M}_{\kappa _{[ 2]}}(z|\zeta _{[ 2]},\zeta _{[ 2]})_
{3}\bar{f}\bar{g},\;\mathcal{L} _{23}(z|f,\III )=\mathcal{M}_
{\kappa _{[ 2]}}(z|w_{[2]},\zeta _{[ 2]})_
{3}\bar{f},\;\mathcal{L}_{23}(z|\III ,\III )=0. \end{eqnarray*}

The condition $[\reve{f},\reve{g}]_{\ast }\in \reve D_{n_
+}^{2k+1}[[\p]]$ implies that $ \reve{M}_{23}(\reve{f},\reve{g})\in
\reve D_{n_+}^{2k+1}[[\p]]$.

In $\hbar ^{6}$-order, Eq. (\ref{2.3.1}) gives four equations.
{}First two of them are \begin{equation} d_
{1}^{\mathrm{ad}}\tilde{M}_{13}(z|f,g)=0,\;\tilde{M}_{13}(z|f)=M_
{23}^{ \prime }(z|f,\III )=M_{23}(z|f,\III )-\mathcal{M}_{\kappa _
{[ 2]}}(z|w_{[2]},\zeta _{[ 2]})_{3}\bar{f}, \label{2.3.3.1}
\end{equation} \begin{equation} m_{23}(\left\{ f,g\right\} ,\III
)=\bar{f}\bar{M}_{23}(|g,\III )-(-1)^{ \varepsilon (f)\varepsilon
(g)}\bar{g}\bar{M}_{23}(|f,\III ). \label{2.3.3.2} \end{equation}
It follows from Eq. (\ref{2.3.3.1}) (with the properties $ M_
{23}(z|f,\III )\in D^{2k+1}_{n_+}[[\p]]$, $\varepsilon _{M_
{23}^{\prime }}=0$, taken into account) \begin{eqnarray*} &&M_
{23}(z|f,\III )=\mathcal{M}_{\kappa _{[ 2]}}(z|w_{[2]},\zeta _{[
2]})_{3}\bar{f}+t_{3}^{0}\mathcal{E}_{z}f(z)+\{t_{3}(z),f(z)\}.
\end{eqnarray*} Eq. (\ref{2.3.3.2}) transforms to \begin{equation*}
m_{23}(\left\{ f,g\right\} ,\III )=(2+n_{+}-n_{-})t_
{3}^{0}]\bar{f}\bar{g }, \end{equation*} and gives \begin{equation*}
t_{3}^{0}=0,\;m_{23}(f,\III )=m_{23}\bar{f}=0. \end{equation*}
Now the third equation takes the form \begin{eqnarray} &&d_
{2}^{\mathrm{ad}}M_{23}^{\prime }(z|f,g,h)=0, \label{2.3.3.3} \\
&&M_{23}^{\prime }(z|f,g,h)=M_{23}(z|f,g,h)-[\mathcal{M}_{\kappa
}(z|\zeta _{[ 2]},\zeta _{[ 2]})_{3}+t_{3}(z)]\bar{f}\bar{g}
\notag \end{eqnarray}

So, we have \begin{eqnarray*} &&M_{23}(z|f,g)=\frac{2\kappa _
{1}\kappa _{3}+\kappa_2^{2}}{6} m_{2|1}(z|f,g)+m_{2|\zeta _
{3}}(z|f,g)+ \\ &&+[\mathcal{M}_{\kappa _{[ 2]}}(z|\zeta _{[
2]},\zeta _{[ 2]})_{3}+t_{3}(z)+c_{23}]\bar{f}\bar{g}+d_
{1}^{\mathrm{ad}}\zeta _{D3}(z|f,g), \\ &&\mathcal{M}_{\kappa _{[
2]}}(z|\zeta _{[ 2]}, \zeta _{[2]})_{3}+t_{3}(z)+c_{23}\in
D^{2k+1}_{n_+}[[\p]], \\ && \varphi_{D3}(z|f)\in D^{2k+1}_{n_
+}[[\p]],\; \zeta _{3}(z)\in E^{2k+1}_{n_+}[[\p]]/D^{2k+1}_{n_
+}[[\p]]. \end{eqnarray*} The last equation is \begin{equation*} d_
{2}^{\mathrm{tr}}m_{23}^{\prime }(f,g,h)=0,\;m_{23}^{\prime
}(f,g,h)=m_{23}(f,g,h)-[\bar{\varphi}_
{D3}(|f)\bar{g}-(-1)^{\varepsilon (f)\varepsilon (g)}\bar{\varphi}_
{D3}(|g)\bar{f}], \end{equation*} and it implies \begin{equation*} m_
{23}(f,g)=\bar{\varphi}_{D3}(|f)\bar{g}-(-1)^{\varepsilon
(f)\varepsilon (g)} \bar{\varphi}_{D3}(|g)\bar{f}-b_
{23}\bar{f}\bar{g}-\mu _{13}(\left\{ f,g\right\} ). \end{equation*}
Introduce a notation \begin{equation*} t_{3}(z)+c_{23}=t_
{D3}(z)+w_{3}(z),\;t_{D3}(z)\in D^{2k+1}_{n_+},\; w_{3}(z)\in
E^{2k+1}_{n_+}[[\p]]/D^{2k+1}_{n_+}[[\p]]. \end{equation*} Then,
we can write \begin{equation*} \reve{M}_
{23}(\reve{f},\reve{g})=\reve{M}_{23|\mathrm{co}}(\reve{f},
\reve{g})+\reve{d}_{1}^{\mathrm{ad}}\reve{M}_
{1D3}(\reve{f},\reve{g}), \end{equation*} where \begin{eqnarray*}
&&M_{23|\mathrm{co}}(f,g)=\frac{2\kappa _{1}\kappa _{3}+\kappa_
2^{2}}{6} m_{2|1}(z|f,g)+m_{2|\zeta _{3}}(z|f,g)+ \\ &&+\left(
\mathcal{M}_{\kappa _{[ 2]}}(z|\zeta _{[ 2]},\zeta _{[ 2]})_
{3}+w_{3}(z)\right) \bar{f}\bar{g},\;\left( \mathcal{M} _{\kappa _
{[ 2]}}(z|\zeta _{[ 2]},\zeta _{[ 2]})_{3}+w_{3}(z)\right) \in
D^{2k+1}_{n_+}[[\p]], \end{eqnarray*} \begin{equation*} M_
{23|\mathrm{co}}(z|f,\III )=\left\{ w_{3}(z),f(z)\right\}
+\mathcal{M} _{\kappa _{[ 2]}}(z|w_{[2]},\zeta _{[ 2]})_
{3}\bar{f},\;m_{23| \mathrm{co}}(f,g)=m_{23|\mathrm{co}}(f,\III
)=0. \end{equation*} \begin{equation*} M_{1D3}(z|f)=\zeta _
{D3}(z|f),\;M_{13}(z|\III )=t_{D3}(z),\;m_{13}(f)=\mu _
{13}(f),\;m_{13}(\III )=b_{23}. \end{equation*}

Performing the similarity transformation $[\reve{f},\reve{g}]_{\ast
}\rightarrow [ \reve{f},\reve{g}]_{\ast T}$ with $\reve{T}(\reve{f}
)=\reve{f}-\hbar ^{6}\reve{M}_{1D3}(\reve{f})+O(\hbar ^{8})$, we
rewrite $ [\reve{f},\reve{g}]_{\ast }$ in the form \begin{equation*}
[ \reve{f},\reve{g}]_{\ast }=\reve{N}_{2|\kappa _{[ 3]},\zeta _{[
3]},w_{[3]}}(\reve{f},\reve{g})+O(\hbar ^{8}). \end{equation*}

Proceeding in the same way, we finally find that up to similarity
transformation, the general form of the deformation of the
 superalgebra $\reve D_{n_+}^{2k+1}$ is given by \begin{eqnarray*}
[ \reve{f},\reve{g}]_{\ast } &=&\reve{N}_{2|\kappa _{[ \infty ]},
\zeta _{[ \infty ]},w_{[\infty ]}}(\reve{f},\reve{g}), \\ \zeta _
{[ \infty ]}(z),w_{[\infty ]}(z) &\in &E^{2k+1}_{n_+}/D^{2k+1}_
{n_+} , \; [\mathcal{M}_{\kappa _{[ \infty ]}}(z|\zeta _{[ \infty
]},\zeta_{[ \infty ]}) +w_{[\infty ]}(z)]\in D^{2k+1}_{n_
+}[[\p]]. \end{eqnarray*}

\section{Superalgebra $\reve D_{n_+}^{n_+ +4}$} In the case under
consideration we have \begin{eqnarray*} &&C(f,g)=\int du\left(
[\mathcal{E}_{u}f(u)]g(u)-f(u)\mathcal{E} _{u}g(u)\right) = \\
&&\,=2\int du[\mathcal{E}_{u}f(u)]g(u),\;\mathcal{E}_
{u}=1-\frac{1}{2} u^{A}\partial _{A}^{u}. \end{eqnarray*}

\subsection{Second adjoint cohomology}

We solve in this section the cohomology equation \begin{equation}
\reve{d}_{2}^{\mathrm{ad}}\reve{M}_
{2}(\reve{f},\reve{g},\reve{h})=0. \label{3.1.0} \end{equation}

It follows from (\ref{1.3}) that $ d_{1}^{\mathrm{ad}}\tilde{M}_
{1}(z|f,g)=0 $. The general solution of this equation has the form $$
M_{2}(z|f,\III )=t^{0}\mathcal{E}_{z}f(z)+\left\{ t^{\prime
}(z),f(z)\right\} , $$ where $$ t^{\prime }(z)\in E^{n_++4}_{n_
+},\;\varepsilon (t^{0})=\varepsilon (t^{\prime }(z))=\varepsilon _
{M_{2}}. $$ Since $ C(f,\mathcal{E}g))-(-1)^{\varepsilon
(f)\varepsilon (g)}C(g,\mathcal{E} f))=0, $ we have $$C(M_
{2}(|f,\III ),g)-(-1)^{\varepsilon (f)\varepsilon (g)}C(M_
{2}(|g,\III ),f)=C(t^{\prime},\{f,g\}) $$

It follows from (\ref{1.3}) also that \begin{equation*} m_
{2}(\left\{ f,g\right\} ,\III )=C(t^{\prime },\{f,g\})
\end{equation*} which implies \begin{equation*} m_{2}(f,\III
)=C(t^{\prime },f)+m\bar{f}. \end{equation*}

{}Further, it follows from (\ref{1.2}) that \be d_
{2}^{\mathrm{ad}}M_{2}^{\prime }(z|f,g,h) = -t^{0}[C(f,g)\mathcal{E}
_{z}h(z)-(-1)^{\varepsilon (f)\varepsilon (g)}C(f,h)\mathcal{E} _
{z}g(z)+C(g,h)\mathcal{E}_{z}f(z)], \label{3.1.1} \ee where
\begin{eqnarray*} &&\,M_{2}^{\prime }(z|f,g,h) =M_
{2}(z|f,g,h)-\tilde{C}_{2}(z|f,g),\\ &&\,\tilde{C}_
{2}(z|f,g)=t^{\prime }(z)C(f,g). \end{eqnarray*} Here we used the
relation \begin{eqnarray*} &&M_{2}(z|h,\III
)C(f,g)-(-1)^{\varepsilon (f)\varepsilon (g)}M_{2}(z|g,\III
)C(f,h)+M_{2}(z|f,\III )C(g,h)= \\ &&\,=t^{0}[\mathcal{E}_
{z}h(z)C(f,g)-(-1)^{\varepsilon (f)\varepsilon (g)} \mathcal{E}_
{z}g(z)C(f,h)+\mathcal{E}_{z}f(z)C(g,h)]-d_{2}^{\mathrm{ad}}
\tilde{C}_{2}(z|f,g,h). \end{eqnarray*}

To solve Eq. (\ref{3.1.1}), consider it in the following domains

1) \begin{equation*} z\bigcap \left[ \mathrm{supp}(f)\bigcup
\mathrm{supp}(g)\bigcup \mathrm{supp} (h)\right]
=\mathrm{supp}(f)\bigcap \left[ \mathrm{supp}(g)\bigcup \mathrm{
supp}(h)\right] =\varnothing . \end{equation*}

2) \begin{equation*} z\bigcap \left[ \mathrm{supp}(f)\bigcup
\mathrm{supp}(g)\right] =\mathrm{supp }(f)\bigcap \left[
\mathrm{supp}(g)\bigcup \mathrm{supp}(h)\right] =\mathrm{
supp}(g)\bigcap \mathrm{supp}(h)=\varnothing . \end{equation*}

3) \begin{equation*} z\bigcap \left[ \mathrm{supp}(f)\bigcup
\mathrm{supp}(g)\bigcup \mathrm{supp} (h)\right] =\varnothing .
\end{equation*}

4) \begin{equation*} \left[ z\bigcup \mathrm{supp}(f)\bigcup
\mathrm{supp}(g)\right] \bigcap \mathrm{supp}(h)=\varnothing .
\end{equation*} In each of these domains, the r.h.s. of Eq.
(\ref{3.1.1}) equals to zero, and we have (for details see
\cite{n>2}) \begin{eqnarray*} &&M_{2}^{\prime }(z|f,g)=\int
dua(u)[\mathcal{E}_{z}f(z)g(u)-f(u)\mathcal{E} _{z}g(z)]+ \\
&&+\int du\left( \{m^{1}(z|u),f(z)\}g(u)-(-1)^{\varepsilon
(f)\varepsilon (g)}\{m^{1}(z|u),g(z)\}f(u)\!\right) + \\ &&+\mu
(z)\int du\left( [\mathcal{E}_{u}f(u)]g(u)-f(u)\mathcal{E} _
{u}g(u)\right) +d_{1}^{\mathrm{ad}}\zeta ^{2}(z|f,g)+M_
{2|\mathrm{loc} }(z|f,g). \end{eqnarray*}

Let \begin{equation*} \left[ z\bigcup \mathrm{supp}(h)\right] \bigcap
\left[ \mathrm{supp} (f)\bigcup \mathrm{supp}(g)\right] =\varnothing
\end{equation*} Then \begin{equation*}
(-1)^{\varepsilon(f)\varepsilon(h)}
(-1)^{\varepsilon(g)\varepsilon(h)}
(-1)^{\varepsilon(h)\varepsilon(M_2)} \{h(z),M_
{2|2}(z|f,g)\}-\hat{M} _{2|2}(z|\{f,g\},h)=0, \end{equation*} which
implies \begin{eqnarray*} &&\{h(z),\mu (z)\}\int du\left(
[\mathcal{E}_{u}f(u)]g(u)-f(u)\mathcal{E} _{u}g(u)\right)
+\mathcal{E}_{z}h(z)\int dua(u)\{f(u),g(u)\}+ \\ &&\,+\int
du\{\hat{m}^{1}(z|u),h(z)\}\{f(u),g(u)\}=t^{0}C(f,g)\mathcal{E} _
{z}h(z). \end{eqnarray*}

Choosing $h(z)=\mathrm{const}$ we obtain \begin{equation*} \int
dua(u)\{f(u),g(u)\}=-t^{0}C(f,g)\;\Longrightarrow t^{0}=0.
\end{equation*}

So, we have \begin{eqnarray*} M_{2}(z|f,g) &=&c_{1}m_
{2|1}(z|f,g)+c_{3}m_{2|3}(z|f,g)+t_{D}(z)C(f,g)+m_{2|\zeta
}(z|f,g)+d_{1}^{\mathrm{ad}}\varphi_{D}(z|f,g), \\ t_{D}(z)
&=&t^{\prime }(z)+c_{4}\in D^{n_++4}_{n_+},\;\varepsilon (t_
{D}(z))=\varepsilon _{M_{2}}, \\ \zeta (z) &\in &E^{n_++4}_{n_
+}/D^{n_++4}_{n_+} ,\;\varepsilon (\zeta (z))=\varepsilon _{M_
{2}},\;\varepsilon _{\varphi_{D}}=\varepsilon _{M_{2}}, \\ M_
{2}(z|f,\III ) &=&\left\{ t_{D}(z),f(z)\right\} , \\ m_{2}(f,\III )
&=&C(t_D,f)+m\bar{f}, \end{eqnarray*} where the bilinear forms $m_
{2|1}$, $m_{2|3}$ and $m_{2|\zeta}$ are the same as in the
Subsection \ref{sectionodd}.

To specify the parameters, use the relations \begin{eqnarray*}
&&(-1)^{\varepsilon (g)\varepsilon (h)}C(m_{2|1}(|f,h),g)-C(m_
{2|1}(|f,g),h)-(-)^{\varepsilon (f)\varepsilon (g)+\varepsilon
(f)\varepsilon (h)}C(m_{2|1}(|g,h),f)= \\ &&\,=-4\int du[f(u)\left(
\overleftarrow{\partial }_{A}\omega ^{AB}\partial _{B}\right)
^{3}g(u)]h(u), \end{eqnarray*} \begin{equation} (-1)^{\varepsilon
(g)\varepsilon (h)}C(m_{2|3}(|f,h),g)-C(m_
{2|3}(|f,g),h)-(-)^{\varepsilon (f)\varepsilon (g)+\varepsilon
(f)\varepsilon (h)}C(m_{2|3}(|g,h),f)=0, \label{3.1.1a}
\end{equation} \begin{eqnarray*} &&(-1)^{\varepsilon (g)\varepsilon
(h)}C(d_{1}^{\mathrm{ad}} \varphi(|f,h),g)-C(d_
{1}^{\mathrm{ad}}\varphi (|f,g),h)-(-)^{\varepsilon (f)\varepsilon
(g)+\varepsilon (f)\varepsilon (h)}C(d_{1}^{\mathrm{ad}}
\varphi(|g,h),f)= \\ &&\,=d_{2}^{\mathrm{tr}}[D_{\zeta }+C_
{\varphi_{D}}](f,g,h), \end{eqnarray*} where \begin{eqnarray*} &&D_
{\zeta }(f,g)=C(\zeta ,g)\bar{f}-(-1)^{\varepsilon (f)\varepsilon
(g)}C(\zeta ,f)\bar{g}, \\ &&C_{\varphi_{D}}(f,g)=C(\varphi_
{D}(|f),g)-(-1)^{\varepsilon (f)\varepsilon (g)}C(\varphi_
{D}(|g),f). \end{eqnarray*} To obtain these relations, we used the
following ones: \begin{eqnarray*} &&\int dzf(z)\mathcal{E}_
{z}g(z)=-\int dz[\mathcal{E}_{z}f(z)]g(z),\; \;n_{-}=n_{+}+4,\\
&&C(\{t,f\},g)-(-1)^{\varepsilon (f)\varepsilon
(g)}C(\{t,g\},f)=C(t,\{f,g\}) \end{eqnarray*} \begin{equation*} \int
dzf(z)[g(z)\left( \overleftarrow{\partial }_{A}\omega ^{AB}\partial
_{B}\right) ^{p}h(z)]=\int dz[f(z)\left( \overleftarrow{\partial }_
{A}\omega ^{AB}\partial _{B}\right) ^{p}g(z)]h(z),\;p=0,1,...,
\end{equation*} \begin{eqnarray*} &&\!\!\mathcal{E}_{z}[f(z)\left(
\overleftarrow{\partial }_{A}\omega ^{AB}\partial _{B}\right)
^{3}g(z)]-[\mathcal{E}_{z}f(z)]\left( \overleftarrow{\partial }_
{A}\omega ^{AB}\partial _{B}\right) ^{3}g(z)-f(z)\left(
\overleftarrow{\partial }_{A}\omega ^{AB}\partial _{B}\right)
^{3}\mathcal{E}_{z}g(z)= \\ &&\!\!\,=2f(z)\left(
\overleftarrow{\partial }_{A}\omega ^{AB}\partial _{B}\right)
^{3}g(z). \end{eqnarray*}

It follows from (\ref{1.2}) also that \begin{eqnarray*} &&d_
{2}^{\mathrm{tr}}m_{2}^{\prime }(f,g,h)=m[C(g,h)\bar{f}
-(-1)^{\varepsilon (f)\varepsilon (g)}C(f,h)\bar{g}+C(f,g)\bar{h}]-
\\ &&-4c_{1}\int dz[f(z)\left( \overleftarrow{\partial }_{A}\omega
^{AB}\partial _{B}\right) ^{3}g(z)]h(z), \\ &&m_{2}^{\prime
}(f,g)=m_{2}(f,g)-[D_{\zeta }+C_{\varphi_{D}}](f,g).
\end{eqnarray*}

Let now \begin{equation*} \left[ \mathrm{supp}(f)\bigcup
\mathrm{supp}(g)\right] \bigcap \mathrm{supp} (h)=\varnothing
,\;g(z)=1,\;z\in \mathrm{supp}(f). \end{equation*} Then we have $ m
C(f,g)\bar{h}=-d_{2}^{\mathrm{tr}}\hat{m}_{2}^{\prime }(f,g,h)$ and
so $m=0$. In such a way \begin{equation*} d_{2}^{\mathrm{tr}}m_
{2}^{\prime }(f,g,h)=-4c_{1}\int dz[f(z)\left(
\overleftarrow{\partial }_{A}\omega ^{AB}\partial _{B}\right)
^{3}g(z)]h(z). \end{equation*} Due to non-triviality of the cocycle
$ \int dz[f(z)\left( \overleftarrow{\partial }_{A}\omega
^{AB}\partial _{B}\right) ^{3}g(z)]h(z) \in C_3(D_{n_+}^{n_-},
\K) $ (see \cite{n>2}) we have $c_1=0$.

So, we obtain \begin{equation*} m_{2}(f,g)=D_{\zeta }(f,g)+C_
{\varphi_{D}}(f,g)-b C(f,g)-d_{1}^{\mathrm{tr} }\mu _{1}(f,g).
\end{equation*} {}Finally, the general solution of Eq. (\ref{3.1.0})
is \begin{eqnarray*} \reve{M}_{2}(\reve{f},\reve{g}) &=&\reve{M}_
{2|1}(\reve{f},\reve{g})+ \reve{d}_{1}^{\mathrm{ad}}\reve{M}_
{1D}(f,g), \\ M_{2|1}(f,g) &=&c_{3}m_{2|3}(z|f,g)+m_{2|\zeta
}(z|f,g),\;m_{2|1}(f,g)=D_{\zeta }(f,g), \\ \reve{M}_
{2|1}(\reve{f},\III ) &=&0, \\ M_{1D}(z|f) &=&\varphi_{D}(z|f),\;M_
{1}(z|\III )=-t_{D}(z),\;m_{1}(f)=\mu _{1}(f),\;m_{1}(\III )=b.
\end{eqnarray*}

\subsubsection{Independence and non-triviality}

Let us suppose that \begin{equation*} \reve{M}_
{2|1}(\reve{f},\reve{g})=\reve{d}_{1}^{\mathrm{ad}}\reve{M} _
{1}(f,g). \end{equation*} It follows from this equation that
\begin{equation} c_{3}m_{2|3}(z|f,g)=d_{1}^{\mathrm{ad}}M_
{1}^{\prime }(z|f,g)-C(f,g)M_{1}(\III ) \label{3.1.2}
\end{equation} for some $M_{1}^{\prime }(z|f)$. Let
\begin{equation*} z\bigcap \left[ \mathrm{supp}(f)\bigcup
\mathrm{supp}(g)\right] =\varnothing ,\;g(z)=1,\;z\in
\mathrm{supp}(f). \end{equation*} It follows from Eq. (\ref{3.1.2})
\begin{equation*} \hat{M}_{1}^{\prime }(z|\{f,g\})+C(f,g)M_{1}(\III
)=0\;\Longrightarrow M_{1}(\III )=0,\; \end{equation*} but then Eq.
(\ref{3.1.2})\ has solutions for $c_{3}=0$ only.

\subsection{Deformations of Lie superalgebra $\reve D_{n_+}^{n_
++4}$}

In this section, we find the general form of the deformation of Lie
superalgebra $\reve D_{n_+}^{n_++4}$, $ [f,g]_{\ast }$,
\begin{eqnarray*} &[\reve{f},\reve{g}]_{\ast }=&M_{2}^{\ast
}(z|\reve{f},\reve{g} )+m_{2}^{\ast }(\reve{f},\reve{g})\III = \sum_
{l=0}^\infty \hbar ^{2l} \reve{M}_{2l}(\reve{f},\reve{g}), \\ &&M_
{2}^{\ast }(z|\reve{f},\reve{g})\in D^{n_++4}_{n_+}[[\p]],\; m_
{2}^{\ast }(\reve{f},\reve{ g})\in \K[[\p]],\;\varepsilon _{M_
{2}^{\ast }}=\varepsilon _{m_{2}^{\ast }}=0, \end{eqnarray*}
satisfying the Jacoby identity.

\subsubsection{$\hbar ^{0}$-order}

We have \begin{equation*} M_{20}(f,g)=\left\{ f(z),g(z)\right\}
,\;m_{20}(f,g)=C(f,g),\;M_{20}(\reve{f },\III )=m_
{20}(\reve{f},\III )=0. \end{equation*}

\subsubsection{$\hbar ^{2}$-order}

In $\hbar ^{2}$-order, Jacoby identity gives the equation
\begin{equation*} \reve{d}_{2}^{\mathrm{ad}}\reve{M}_{21}(f,g,h)=0
\end{equation*} The general solution of this equation is found in the
preceding section \begin{equation*} \reve{M}_
{21}(\reve{f},\reve{g})=\reve{M}_{2|11}(\reve{f},\reve{g})+
\reve{d}_{1}^{\mathrm{ad}}\reve{M}_{1D1}(f,g), \end{equation*}
\begin{eqnarray*} &&M_{2|11}(f,g)=c_{3}m_{2|3}(z|f,g)+m_{2|\zeta
_{1}}(z|f,g),\;\varepsilon (\zeta _{1})=0, \\ &&m_{2|11}(f,g)=D_
{\zeta _{1}}(f,g), \\ &&\reve{M}_{2|11}(\reve{f},\III )=0 \\
&&\reve{M}_{1D1}(f)=\varphi_{D1}(z|f)+m_{11}(f)\III ,\;\reve{M} _
{1D1}(\III )=M_{1D1}(z|\III )+b_{1}\III . \end{eqnarray*}

\subsubsection{Higher orders}

Introduce a form \begin{eqnarray*} \reve{S}_{2|\zeta ,c_
{3}}(\reve{f},\reve{g}) &=&S_{2|\zeta ,c_{3}}(z|
\reve{f},\reve{g})+s_{2|\zeta ,c_{3}}(\reve{f},\reve{g})\III , \\
S_{2|\zeta ,c_{3}}(z|f,g) &=&\left\{ f(z),g(z)\right\} +c_{3}m_
{2|3}(z|f,g)+m_{2|\zeta }(z|f,g), \\ s_{2|\zeta,c_
{3}}(f,g)=C(f,g)+D_{\zeta }(f,g), \\ S_{2|\zeta ,c_
{3}}(\reve{f},\III ) &=&s_{2|\zeta ,c_{3}}(\reve{f},\III )=0.
\end{eqnarray*} The form $\reve{S}_{2|\zeta ,c_{3}}(z|\reve{f},
\reve{g})$ satisfy the Jacoby identity \begin{equation*}
(-1)^{\varepsilon (\reve{f})\varepsilon (\reve{h})}\reve{S}_{2|\zeta
,c_{3}}(\reve{S}_{2|\zeta ,c_
{3}}(\reve{f},\reve{g}),\reve{h})+\mathrm{
cycle}(\reve{f},\reve{g},\reve{h})=0. \end{equation*} To prove this
fact, one should use the fact that the form $S_{2|\zeta ,c_
{3}}(z|f,g)$ satisfy the Jacoby identity (see \cite{deform}), the
relation (\ref{3.1.1a}), and relations \begin{equation*}
(-1)^{\varepsilon (f)\varepsilon (h)}C(\{f,g\},h)+\mathrm{cycle}
(f,g,h)=(-1)^{\varepsilon (f)\varepsilon (h)}d_
{2}^{\mathrm{tr}}C(f,g,h)=0, \end{equation*} \begin{eqnarray*}
&&\,(-1)^{\varepsilon (f)\varepsilon (h)}C(m_{2|\zeta
}(|f,g),h)+\mathrm{ cycle}(f,g,h)=-(-1)^{\varepsilon (f)\varepsilon
(h)}d_{2}^{\mathrm{tr} }D_{\zeta }(f,g,h), \\ &&\,(-1)^{\varepsilon
(f)\varepsilon (h)}D_{\zeta }(\{f,g\},h)+\mathrm{cycle}
(f,g,h)=(-1)^{\varepsilon (f)\varepsilon (h)}d_{2}^{\mathrm{tr}}D_
{\zeta }f,g,h), \\ &&\,(-1)^{\varepsilon (f)\varepsilon (h)}C(m_
{2|\zeta }(|f,g),h)+(-1)^{\varepsilon (f)\varepsilon (h)}D_{\zeta
}(\{f,g\},h)+ \mathrm{cycle}(f,g,h)=0, \end{eqnarray*}
\begin{eqnarray*} (-1)^{\varepsilon (f)\varepsilon (h)}D_{\zeta }(m_
{2|3}(|f,g),h)+\mathrm{ cycle}(f,g,h) &=&0, \\ (-1)^{\varepsilon
(f)\varepsilon (h)}D_{\zeta }(m_{2|\zeta }(|f,g),h)+
\mathrm{cycle}(f,g,h) &=&0. \end{eqnarray*}

Performing the similarity transformation $[\reve{f},\reve{g}]_{\ast
}\rightarrow [ \reve{f},\reve{g}]_{\ast T}$ with $\reve{T}(\reve{f}
)=\reve{f}-\hbar ^{2}\reve{M}_{11}(\reve{f})+O(\hbar ^{4})$, one can
rewrite $[\reve{f},\reve{g}]_{\ast }$ in the form \begin{equation*}
[ \reve{f},\reve{g}]_{\ast }=\reve{S}_{2|\zeta _{[ 1]},c_
{3[1]}}(\reve{f},\reve{g})+\hbar ^{4}\reve{M}_{22}(\reve{f},\reve{
g})+O(\hbar ^{6}). \end{equation*}

In $\hbar ^{4}$-order, the Jacoby identity gives the equation
\begin{equation*} \reve{d}_{2}\reve{M}_{22}(\reve{f},\reve{g})=0.
\end{equation*}

Proceeding in the same way, we finally find that up to a similarity
transformation, the general form of the deformation of the centrally
extended super Poincare algebra is given by \begin{equation*} [
\reve{f},\reve{g}]_{\ast }=\reve{S}_{2|\zeta _{[ \infty ]},c_
{3[\infty ]}}(\reve{f},\reve{g}). \end{equation*}

%%%%%%%%%%%%%%%%%%%%%%%%%%%%%%%%%%%%%%%%%%%%%%%%%%%%%%%%%%%%%%%%


\begin{thebibliography}{99}


\bibitem{Schei97}
{\it M.Scheunert, R.B.Zhang}, J.Math.Phys. {\bf 39} (1998)
5024-5061, q-alg/9701037.

\bibitem{n>2}
{\it S.Konstein, A.Smirnov and I.Tyutin}, Cohomologies of the Poisson
Superalgrbra,\\ hep-th/0312109;
\\
{\it S.Konstein, A.Smirnov and I.Tyutin}, submitted to TM{}F.

\bibitem{deform}
{\it S.Konstein, A.Smirnov and I.Tyutin},
General form of deformation of Poisson superbracket, hep-th/0401023.

\bibitem{n=2} {\it S.Konstein and I.Tyutin},
Cohomologies of the Poisson superalgebra
on \\ (2,n)-superdimensional spaces, hep-th/0411235.

\end{thebibliography}
\end{document}